\documentclass[11pt]{article}

\usepackage[a4paper,margin=1in]{geometry}
\usepackage{amsmath,amssymb,amsfonts,mathtools}
\usepackage{physics}
\usepackage{bm}
\usepackage{subfig}
\usepackage[colorlinks=true,linkcolor=blue,citecolor=blue,urlcolor=blue]{hyperref}

\numberwithin{equation}{section}

\newcommand{\cC}{\mathcal C}
\newcommand{\cH}{\mathbb H}
\newcommand{\cT}{\mathcal T}
\newcommand{\slit}{\mathrm{slit}}
\newcommand{\cR}{\mathcal R}
\newcommand{\cD}{\mathbb D}
\newcommand{\Hf}{\operatorname{Hf}}
\newcommand{\ii}{\mathrm i}

\title{Excited states entanglement after measurements in a CFT}
\begin{document}
\begin{titlepage}
	\title{\textbf {Entanglement of excited states after measurements in conformal field theory}}
    \author{Hui-Huang Chen$^{a}$\footnote{chenhh@jxnu.edu.cn}}
	\maketitle
	\underline{}
	\vspace{-12mm}
	\begin{center}
		{\it
             $^a$School of Physics, Jiangxi Normal University,\\ Nanchang 330022, China\\
		}
		\vspace{10mm}
	\end{center}
\begin{abstract}

We study the entanglement of low-energy excited states after a fixed-outcome projective measurement on a spatial interval in a (1+1)-dimensional conformal field theory (CFT). We restrict to post-selected measurement outcomes whose induced boundary condition flows in the infrared to a conformal boundary condition. Such an outcome is represented by a slit carrying that boundary condition, while excited states are introduced by operator insertions in the Euclidean path integral. After mapping the replicated slit geometry to a disk, the excited-state contribution to the post-measurement R\'enyi entropy is expressed as a normalized boundary-CFT correlation function.

We apply this framework to the compact free boson CFT. For the chiral current excitation, the relevant ratios are given by current hafnians. We also study coherent superpositions of \(J\) and \(\bar J\), and of conjugate compact vertex operators. In the conjugate-vertex case, the ordinary-cylinder second Rényi ratio is independent of the relative phase, whereas the finite-slit post-measurement ratio contains phase-sensitive interference terms. Finally, we describe free-fermion and multi-Slater determinant methods for testing these predictions in the critical XX chain.
\end{abstract}
\end{titlepage}
\thispagestyle{empty}
\newpage
\tableofcontents
\newpage

\section{Introduction}
\label{sec:introduction}

Entanglement is a universal probe of low-energy structure in quantum many-body
systems. In one spatial dimension, the ground-state entanglement of a critical
system is governed by the conformal field theory (CFT) describing its infrared
fixed point. For a periodic system of length \(L\), the ground-state R\'enyi entropy of an
interval of length \(\ell\) takes the following universal form \cite{Holzhey1994,CalabreseCardy2004,
CalabreseCardy2009}
\begin{equation}
    S_n(\ell)
    =
    \frac{c}{6}
    \left(
        1+\frac{1}{n}
    \right)
    \log
    \left[
        \frac{L}{\pi \epsilon}
        \sin\frac{\pi\ell}{L}
    \right]
    +
    \gamma_n ,
    \label{eq:intro-ground-renyi}
\end{equation}
where \(c\) is the central charge, \(\epsilon\) is a short-distance cutoff, and
\(\gamma_n\) is nonuniversal. The (1+1)-dimensional CFT framework provides the natural
language for branch-point twist fields, boundary conditions, and conformal
maps \cite{DiFrancesco1997,Cardy1984,Cardy2004BCFT}, all of which play a central role in studying quantum information aspects of one-dimensional critical systems.

Low-energy excited states contain additional universal entanglement information. If an excited state is created by a scaling operator \(\Upsilon\), the excess R\'enyi entropy relative to the ground state can be expressed as a normalized \(2n\)-point function of \(\Upsilon\) and \(\Upsilon^\dagger\) on the replicated geometry \cite{AlcarazBerganzaSierra2011, BerganzaAlcarazSierra2012}. This construction has been tested in critical spin chains and gives a direct bridge between operator content in CFT and entanglement properties of low-energy lattice excitations.

Measurements provide another way of reorganizing entanglement. In monitored dynamics, the competition between unitary evolution and local measurements can drive measurement-induced phase transitions between distinct entanglement phases \cite{LiChenFisher2018,SkinnerRuhmanNahum2019}. Replica mappings to statistical mechanics and conformal-invariance arguments have played an important role in understanding these transitions \cite{JianYouVasseurLudwig2020,BaoChoiAltman2020}. In particular, numerical and analytical studies have provided evidence that measurement-induced
critical points in hybrid circuits are described by nonunitary CFTs, with universal data accessible through entanglement, mutual information, operator
dimensions, and boundary spectra \cite{LiChenLudwigFisher2021, ZabaloGullansWilsonVasseurLudwigGopalakrishnanHusePixley2022, KumarAzizChakrabortyLudwigGopalakrishnanPixleyVasseur2024}.

A related but static problem asks how measurements performed on a many-body state induce entanglement between the unmeasured degrees of freedom. This
point of view goes back to localizable entanglement \cite{VerstraetePoppCirac2004} and has recently been developed as a probe of wavefunction structure and quantum criticality \cite{LinYeZouSangHsieh2023,ChengWenGopalakrishnanVasseurPotter2024}. For critical one-dimensional systems, a particularly tractable setting is a fixed local measurement outcome that flows, in the infrared, to a conformal boundary condition. In the Euclidean path integral, the measured region is then represented as a slit carrying that boundary condition. The post-measurement entanglement can be computed using boundary CFT and conformal maps from slit geometries to the upper half-plane, annuli, or finite cylinders \cite{Rajabpour2015,Rajabpour2016,NajafiRajabpour2016}. This approach covers not only a single measured interval but also more general geometries, including open systems, finite-temperature setups, and multiple measured intervals, where the CFT problem is related to Casimir energies of floating objects \cite{NajafiRajabpour2016}.  More recent work has studied finite-density and weak measurements in critical chains, where measurements act as defects in the Euclidean CFT and can lead to effective central charges, marginal post-selected defects, or measurement-altered critical states
\cite{WeinsteinSajithAltmanGarratt2023,YangMaoJian2023,GarrattWeinsteinAltman2023,MurcianoSalaLiu2023}.

In this paper, we consider a critical one-dimensional system on a circle and perform a projective measurement on a spatial interval. Throughout this work we restrict to post-selected measurement outcomes whose measurement-induced boundary perturbation flows in the infrared to a conformal
boundary condition \(\mathfrak a\). Generic measurement records need not satisfy this assumption. Measurement-induced boundary RG flows and boundary transitions in critical systems have been studied, for example, in Ref.~\cite{LiuMurcianoMrossAlicea2025}. The measured interval becomes a slit, and the unmeasured system is described by a CFT on a slit cylinder. We then ask how low-energy excited states modify the post-measurement entanglement between subregions of the unmeasured system. Excited-state R\'enyi entropies in systems with conformal boundaries have also been studied directly in boundary conformal field theory (BCFT) \cite{TaddiaOrtolaniPalmai2016}. The normalized boundary correlation function structure is closely related to the one used below. Here, however, the conformal boundary is generated by a finite measurement slit rather than being a pre-existing spatial boundary.

For a state created by a single scaling operator \(\Upsilon\), the resulting fixed-outcome R\'enyi entropy has the universal structure
\begin{equation}
    S^{(\Upsilon,M)}_n
    =
    S^{(0,M)}_n
    +
    \frac{1}{1-n}
    \log F^{(n)}_{\Upsilon,\mathfrak a}.
    \label{eq:intro-main-formula}
\end{equation}
Here \(S^{(0,M)}_n\) is the post-measurement ground-state R\'enyi entropy fixed by the slit geometry, while \(F^{(n)}_{\Upsilon,\mathfrak a}\) is a
normalized operator correlation function on the replicated slit-cylinder
geometry. After mapping the slit cylinder to the upper half-plane and
uniformizing the replicated surface, this ratio becomes a correlation
function on a disk with boundary condition \(\mathfrak a\). For coherent
superpositions of scaling states, the same geometry applies, but
\(F^{(n)}_{\Upsilon,\mathfrak a}\) becomes a finite sum of such multi-point functions, with
selection rules determined by the CFT operator algebra.

We work out the framework in the compact free boson CFT through three examples: 
the chiral current \(J=\ii\partial\phi\), a coherent \(J/\bar J\) superposition, 
and a coherent superposition of conjugate compact vertex operators. The current 
example gives closed hafnian expressions for the second and third R\'enyi ratios. 
The two superposition examples show how the measurement-induced boundary
produces relative-phase-dependent interference between different sectors in
post-measurement entanglement. For the conjugate-vertex superposition, the
ordinary-cylinder second R\'enyi ratio is independent of the relative phase,
while the finite-slit result contains explicit \(\cos\theta\) and
\(\cos2\theta\) contributions. This provides a simple example in which a
fixed-outcome measurement activates phase sensitivity that is absent in the
unmeasured case.

We also describe lattice checks in the critical XX chain. For the current
excitation, both the ground state and the particle-hole excited state are
Slater determinants, so fixed-outcome post-measurement R\'enyi entropies can
be computed from conditional free-fermion correlation matrices. For the left-right 
current and conjugate-vertex superpositions, the state is a coherent 
sum of two Slater determinants and is not Gaussian. Integer R\'enyi entropies can 
nevertheless be computed efficiently using a multi-Slater determinant method. 
This gives a direct numerical route to testing the
CFT predictions for both types of excitations.

The rest of the paper is organized as follows. In
Sec.~\ref{sec:ground-state}, we introduce the fixed-outcome slit-cylinder
geometry, derive the conformal map to the upper half-plane, and obtain the
post-measurement ground-state R\'enyi entropy. In
Sec.~\ref{sec:excited-replicated-geometry}, we extend this geometry to
operator-created excited states and derive the general normalized
correlation-function formula on the replicated slit cylinder. In
Sec.~\ref{sec:current-excitation}, we specialize to the compact-boson current
and evaluate the corresponding hafnian expressions, including their
unmeasured-cylinder limits. In Sec.~\ref{sec:current-superposition}, we study the 
superposition state of left and right current operators.
In Sec.~\ref{sec:vertex-superposition}, we study
coherent superpositions of conjugate vertex operators and show that the finite-slit geometry activates
a phase dependence that is absent in the ordinary-cylinder limit. In Sec.~\ref{sec:numerics-current}, we describe the
correlation-matrix method for the current excitation, and in
Sec.~\ref{sec:numerics-superposition}, we present the multi-Slater determinant
method for superposition states. We conclude in
Sec.~\ref{sec:conclusion} with a discussion of possible extensions.

\section{Ground-state entanglement entropy after a measurement}\label{sec:ground-state}
In this section, we review the BCFT derivation of the known result for the post-measurement ground-state entanglement entropy.
\subsection{The geometry}\label{sec:setup-geometry}
We consider a CFT on a Euclidean cylinder \(\cC\) with complex coordinate
\(w=\sigma+\ii\tau\), where \(\sigma\sim\sigma+L\). The physical quantum system
lives on the \(\tau=0\) slice. We perform a projective measurement on the
interval \(A=[-s/2,s/2]\), where \(s\) is the length of the measured
region. We restrict throughout to post-selected measurement outcomes whose induced
boundary condition flows in the infrared to a conformal boundary condition \(\mathfrak a\). In the Euclidean
path-integral representation, the measured interval is removed from the time
slice and becomes a slit,
\begin{equation}
    \cC_{\slit}^{\mathfrak a}
    =
    \cC\setminus A,
    \label{eq:slit-cylinder}
\end{equation}
with the two sides of the slit carrying the boundary condition
\(\mathfrak a\). This is the boundary-CFT representation of the post-selected
measurement outcome.

The unmeasured part of the spatial circle is
\(\bar{A}=S^1_L\setminus A\). We choose the subsystem
whose entropy is computed to be the interval adjacent to the right endpoint
of the measured region,
\begin{equation}
    B
    =
    \left[
        \frac{s}{2},\frac{s}{2}+l
    \right]
    \subset \overline{A} .
    \label{eq:subsystem-B}
\end{equation}
Thus one endpoint of \(B\) lies on the measurement-induced boundary, while
the other endpoint is at \(P=s/2+l\). The complement of \(B\) inside the
unmeasured region will be denoted by \(C\). We are interested in
the bipartite entanglement between \(B\) and \(C\) in the
post-measurement state.

Because one endpoint of \(B\) coincides with the boundary created by the
measurement, the replica construction contains only one nontrivial bulk branch
point, located at \(P\). Hence the post-measurement ground-state R\'enyi
entropy is determined by a one-point function of the branch-point twist field
\(\cT_n\) on the slit cylinder. The twist field has weights
\(h_n=\bar h_n=\frac{c}{24}(n-1/n)\), and total scaling dimension
\(\Delta_n=h_n+\bar h_n=\frac{c}{12}(n-1/n)\). Up to a nonuniversal
normalization factor, the replicated partition function is therefore
\begin{equation}
    \Tr \rho_B^n
    \propto
    \left\langle
        \cT_n(P)
    \right\rangle_{\cC_{\slit}^{\mathfrak a}} .
    \label{eq:twist-one-point-slit}
\end{equation}
This is the fixed-outcome analogue of the standard twist-field representation
of R\'enyi entropies in CFT \cite{CalabreseCardy2004,CalabreseCardy2009}.

\subsection{Conformal map to the upper half-plane}

The slit cylinder described above can be mapped to the upper half-plane by
\begin{equation}
    z(w)
    =
    \left[
    \frac{
        \sin\frac{\pi}{2L}(s+2w)
    }{
        \sin\frac{\pi}{2L}(s-2w)
    }
    \right]^{1/2}.
    \label{eq:slit-map}
\end{equation}
The square-root branch is chosen so that the physical domain maps to
\(\cH=\{z:\operatorname{Im}z>0\}\). Under this map, the slit \(A\)
is sent to the real axis of the upper half-plane, where the boundary condition
\(\mathfrak a\) is imposed.

The entangling point \(P=s/2+l\) is mapped to
\begin{equation}
    z_P=z(P)=\ii y,
    \qquad
    y=
    \left[
    \frac{
        \sin\frac{\pi(l+s)}{L}
    }{
        \sin\frac{\pi l}{L}
    }
    \right]^{1/2}.
    \label{eq:zP-y}
\end{equation}
Its reflected image is \(\bar z_P=-\ii y\), which will be useful when applying
the doubling trick in boundary CFT.

The logarithmic derivative of the map is
\begin{equation}
    \partial_w\log z(w)
    =
    \frac{\pi}{2L}
    \left[
        \cot\frac{\pi(s+2w)}{2L}
        +
        \cot\frac{\pi(s-2w)}{2L}
    \right].
    \label{eq:log-derivative}
\end{equation}
Evaluating it at \(w=P\) gives
\begin{equation}
    \left|
    \frac{\partial_w z(P)}{z(P)}
    \right|
    =
    \frac{\pi}{2L}
    \frac{
        \sin\frac{\pi s}{L}
    }{
        \sin\frac{\pi(l+s)}{L}
        \sin\frac{\pi l}{L}
    }.
    \label{eq:derivative-at-P}
\end{equation}
Since \(|z(P)|=y\), this also fixes \(|\partial_w z(P)|\), which enters the
transformation law of the twist-field one-point function.

\subsection{Ground-state entanglement entropy after the fixed measurement}

The one-point function of a bulk primary field of scaling dimension
\(\Delta_n\) in the upper half-plane has the form
\begin{equation}
    \left\langle
        \cT_n(z_P,\bar z_P)
    \right\rangle_{\cH_{\mathfrak a}}
    =
    C_{n,\mathfrak a}
    \left(
        \frac{\epsilon}{|z_P-\bar{z}_P|}
    \right)^{\Delta_n}.
    \label{eq:uhp-one-point}
\end{equation}
Here \(\epsilon\) is a UV cutoff and \(C_{n,\mathfrak a}\) is a nonuniversal
boundary-dependent amplitude. Under the conformal transformation
\(w\mapsto z\), the one-point function on the slit cylinder is
\begin{equation}
    \left\langle
        \cT_n(P)
    \right\rangle_{\cC_{\slit}^{\mathfrak a}}
    =
    \left|
        \partial_w z(P)
    \right|^{\Delta_n}
    \left\langle
        \cT_n(z_P,\bar z_P)
    \right\rangle_{\cH_{\mathfrak a}} .
    \label{eq:cyl-one-point}
\end{equation}
Combining Eqs.~\eqref{eq:zP-y}, \eqref{eq:derivative-at-P}, and
\eqref{eq:uhp-one-point}, one finds that, up to nonuniversal constants,
\begin{equation}
    \left\langle
        \cT_n(P)
    \right\rangle_{\cC_{\slit}^{\mathfrak a}}
    \propto
    \left[
        \frac{\pi \epsilon}{L}
        \frac{
            \sin\frac{\pi s}{L}
        }{
            \sin\frac{\pi(l+s)}{L}
            \sin\frac{\pi l}{L}
        }
    \right]^{\Delta_n}.
    \label{eq:twist-one-point-scaling}
\end{equation}
Numerical factors independent of \(l,s,L\), such as powers of \(2\), have been
absorbed into the nonuniversal normalization.

The post-measurement ground-state R\'enyi entropy is therefore
\begin{equation}
    S^{(0,M)}_n(l,s,L)
    =
    \frac{1}{1-n}
    \log
    \left\langle
        \cT_n(P)
    \right\rangle_{\cC_{\slit}^{\mathfrak a}}
    +
    \text{nonuniversal constant}.
\end{equation}
Using the expression for \(\Delta_n\), this becomes
\begin{equation}
    S^{(0,M)}_n(l,s,L)
    =
    \frac{c}{12}
    \left(
        1+\frac{1}{n}
    \right)
    \log
    \left[
        \frac{L}{\pi \epsilon}
        \frac{
            \sin\frac{\pi(l+s)}{L}
            \sin\frac{\pi l}{L}
        }{
            \sin\frac{\pi s}{L}
        }
    \right]
    +
    \gamma_{n,\mathfrak a}.
    \label{eq:post-measurement-ground-renyi}
\end{equation}
The constant \(\gamma_{n,\mathfrak a}\) absorbs the cutoff convention, the
normalization of the twist operator, the boundary entropy associated with
\(\mathfrak a\), and other nonuniversal terms.

Taking the limit \(n\to1\) gives the von Neumann entropy
\begin{equation}
    S^{(0,M)}_1(l,s,L)
    =
    \frac{c}{6}
    \log
    \left[
        \frac{L}{\pi \epsilon}
        \frac{
            \sin\frac{\pi(l+s)}{L}
            \sin\frac{\pi l}{L}
        }{
            \sin\frac{\pi s}{L}
        }
    \right]
    +
    \gamma_{1,\mathfrak a}.
    \label{eq:post-measurement-ground-vn}
\end{equation}
Equations~\eqref{eq:post-measurement-ground-renyi} and
\eqref{eq:post-measurement-ground-vn} are the fixed-outcome BCFT predictions
for the post-measurement ground-state entanglement in the adjacent-interval
geometry \cite{Rajabpour2015,Rajabpour2016}.

\subsection{Comments on the UV cutoff and the unmeasured limit}
\label{subsec:unmeasured-limit}

The slit geometry is topologically different from the ordinary cylinder, so
one should not set \(s=0\) directly inside the slit-cylinder formula. The
correct way to recover the ordinary unmeasured result is to set the slit
length equal to the microscopic cutoff: \(s=\epsilon\), and then take the continuum scaling limit
\begin{equation}
    \frac{\epsilon}{L}\to0,
    \qquad
    \frac{l}{L}\ \text{fixed}.
\end{equation}
Indeed, using
\begin{equation}
    \sin\frac{\pi(l+\epsilon)}{L}
    =
    \sin\frac{\pi l}{L}
    +
    O\!\left(\frac{\epsilon}{L}\right),
    \qquad
    \sin\frac{\pi \epsilon}{L}
    =
    \frac{\pi \epsilon}{L}
    +
    O\!\left(\frac{\epsilon^3}{L^3}\right),
\end{equation}
Eq.~\eqref{eq:post-measurement-ground-renyi} reduces to
\begin{equation}
    S^{(0,M)}_n(l,\epsilon,L)
    =
    \frac{c}{6}
    \left(
        1+\frac{1}{n}
    \right)
    \log
    \left[
        \frac{L}{\pi \epsilon}
        \sin\frac{\pi l}{L}
    \right]
    +
    \gamma'_n
    +
    O\!\left(\frac{\epsilon}{L}\right),
    \label{eq:ordinary-limit}
\end{equation}
which is the usual finite-size CFT result for the ground-state R\'enyi entropy
of a single interval on a circle \cite{CalabreseCardy2004,CalabreseCardy2009}.
This observation will be useful later when comparing the excited-state
post-measurement formulas with their unmeasured-cylinder limits.
\section{Excited states and the replicated slit geometry}
\label{sec:excited-replicated-geometry}
We now extend the fixed-outcome slit-cylinder construction to low-energy excited states. The measurement outcome is assumed to flow to a conformal boundary condition \(\mathfrak a\), so that the Euclidean path integral is performed on the slit cylinder \(\cC_{\slit}^{\mathfrak a}\). Excited states are represented by operator insertions in the infinite Euclidean past and future.

\subsection{Excited states created by CFT operators}

We first consider an excited state created by a single scaling operator
\(\Upsilon\) of weights \((h,\bar h)\). In cylinder quantization, the ket and
bra are represented by inserting \(\Upsilon\) in the infinite Euclidean past
and \(\Upsilon^\dagger\) in the infinite Euclidean future:
\begin{equation}
    |\Upsilon\rangle
    =
    \lim_{\tau\to-\infty}
    \Upsilon(w,\bar w)|0\rangle,
    \qquad
    \langle\Upsilon|
    =
    \lim_{\tau\to+\infty}
    \langle0|\Upsilon^\dagger(w,\bar w).
    \label{eq:excited-state-cylinder}
\end{equation}
We denote the corresponding insertion points by \(w_-=-\ii\infty\) and
\(w_+=+\ii\infty\). After the fixed projective measurement, the path integral
is performed on the slit cylinder \(\cC_{\slit}^{\mathfrak a}\), and the
normalization of the post-measurement excited-state density matrix is
\begin{equation}
    Z_{\Upsilon,\mathfrak a}
    =
    \left\langle
        \Upsilon(w_-,\bar w_-)
        \Upsilon^\dagger(w_+,\bar w_+)
    \right\rangle_{\cC_{\slit}^{\mathfrak a}} .
    \label{eq:excited-normalization}
\end{equation}

To compute \(\Tr\rho^n_{\Upsilon,B}\), we glue \(n\) copies of the slit
cylinder cyclically along the cut associated with \(B\). The resulting
surface is denoted by \(\cR^{M,\mathfrak a}_n(P)\), where \(P\) is the
non-boundary endpoint of \(B\). Each sheet contains one pair of insertions,
\(\Upsilon(w_-)\) and \(\Upsilon^\dagger(w_+)\). Thus
\begin{equation}
    \Tr\rho^n_{\Upsilon,B}
    =
    F^{(n)}_{\Upsilon,\mathfrak a}(l,s,L)\Tr\rho^n_{0,B}\,,
    \label{eq:rho-factorization}
\end{equation}
with
\begin{equation}
    F^{(n)}_{\Upsilon,\mathfrak a}
    =
    \frac{
    \left\langle
        \prod_{k=0}^{n-1}
        \Upsilon_k(w_-,\bar w_-)
        \Upsilon^\dagger_k(w_+,\bar w_+)
    \right\rangle_{\cR^{M,\mathfrak a}_n(P)}
    }{
    \left[
    \left\langle
        \Upsilon(w_-,\bar w_-)
        \Upsilon^\dagger(w_+,\bar w_+)
    \right\rangle_{\cC_{\slit}^{\mathfrak a}}
    \right]^n
    } .
    \label{eq:F-general-Rn}
\end{equation}
The sheet label is indicated by the subscript \(k\). This gives the
fixed-outcome analogue of the usual excited-state CFT formula on the ordinary
cylinder \cite{AlcarazBerganzaSierra2011,BerganzaAlcarazSierra2012}. The
R\'enyi entropy is therefore
\begin{equation}
    S^{(\Upsilon,M)}_n
    =
    S^{(0,M)}_n
    +
    \frac{1}{1-n}
    \log F^{(n)}_{\Upsilon,\mathfrak a},
    \label{eq:S-excited-general}
\end{equation}
and, provided \(F^{(1)}_{\Upsilon,\mathfrak a}=1\), the von Neumann entropy is
obtained by analytic continuation as
\begin{equation}
    S^{(\Upsilon,M)}_1
    =
    S^{(0,M)}_1
    -
    \left.
    \partial_n
    \log F^{(n)}_{\Upsilon,\mathfrak a}
    \right|_{n=1}.
    \label{eq:S-vn-general}
\end{equation}

\subsection{Uniformization of the replicated slit geometry}\label{sec:slit-geometry}

We now map the replicated slit geometry to a single disk. The slit cylinder is
first mapped to the upper half-plane by the map \(z(w)\) in
Eq.~\eqref{eq:slit-map}. The entangling point \(P=s/2+l\) maps to
\(z_P=\ii y\), with
\begin{equation}
    y=
    \left[
        \frac{\sin\frac{\pi(l+s)}{L}}
             {\sin\frac{\pi l}{L}}
    \right]^{1/2}.
    \label{eq:y-again}
\end{equation}
After this map, the replicated geometry is an \(n\)-sheeted cover of the
upper half-plane branched at \(z_P=\ii y\). By the doubling trick, it can be
viewed as a full-plane cover branched at \(\ii y\) and \(-\ii y\). A convenient
uniformizing coordinate is
\begin{equation}
    \xi_k(z)
    =
    e^{2\pi \ii k/n}
    \left(
        \frac{z-\ii y}{z+\ii y}
    \right)^{1/n},
    \qquad
    k=0,\ldots,n-1.
    \label{eq:xi-k}
\end{equation}
For \(n=1\), this reduces to the disk coordinate
\begin{equation}
    \eta(z)=\frac{z-\ii y}{z+\ii y}.
    \label{eq:eta-def}
\end{equation}
Thus the replicated surface is uniformized to a disk \(\cD_{\mathfrak a}\)
with boundary condition \(\mathfrak a\).

The operator insertions at \(w_\pm=\pm \ii\infty\) first map to
\(z_\pm=e^{\ii(\pi/2\mp\alpha_s)}\), where \(\alpha_s=\pi s/(2L)\). Their images
under the \(n=1\) disk map are
\begin{equation}
    \eta_\pm=\frac{z_\pm-\ii y}{z_\pm+\ii y}.
    \label{eq:eta-pm}
\end{equation}
On the \(n\)-sheeted surface, the corresponding uniformized insertion points
are
\begin{equation}
    \xi_{\pm,k}
    =
    e^{2\pi \ii k/n}\eta_\pm^{1/n},
    \qquad
    k=0,\ldots,n-1.
    \label{eq:xi-pm}
\end{equation}

We will also need the derivative of the uniformizing map at the insertion
points. Since \(\dd\eta/\dd z=2\ii y/(z+\ii y)^2\), one has
\begin{equation}
    \xi'_{\pm,k}
    \equiv
    \left.
    \frac{\dd\xi_k}{\dd z}
    \right|_{z=z_\pm}
    =
    \frac{1}{n}
    e^{2\pi \ii k/n}
    \eta_\pm^{1/n-1}\eta'_\pm,
    \qquad
    \eta'_\pm
    =
    \frac{2\ii y}{(z_\pm+\ii y)^2}.
    \label{eq:xi-prime-pm}
\end{equation}
Equivalently,
\begin{equation}
    \xi'_{\pm,k}
    =
    \frac{2\ii y}{n}
    \frac{\xi_{\pm,k}}{(z_\pm-\ii y)(z_\pm+\ii y)} .
    \label{eq:xi-prime-equiv}
\end{equation}
All dependence on the subsystem geometry is therefore encoded in
\(y\), \(\alpha_s\), and the two disk coordinates \(\eta_\pm\).

\subsection{General primary-field formula on the uniformized disk}

We now transform the operator insertions in
Eq.~\eqref{eq:F-general-Rn} to the uniformized disk. For a primary field
\(\Upsilon\) of weights \((h,\bar h)\), a conformal transformation
\(z\mapsto \xi\) gives
\begin{equation}
    \Upsilon(z,\bar z)
    =
    \left(
        \frac{\dd \xi}{\dd z}
    \right)^h
    \left(
        \frac{\dd \bar\xi}{\dd \bar z}
    \right)^{\bar h}
    \Upsilon(\xi,\bar\xi).
    \label{eq:primary-transform}
\end{equation}
The Jacobian factors from the first map \(w\mapsto z\) cancel between the
numerator and denominator of the normalized ratio
\(F^{(n)}_{\Upsilon,\mathfrak a}\). Thus the only remaining universal
Jacobian factors are those associated with the uniformization
\(z\mapsto \xi\).

The result is
\begin{equation}
    F^{(n)}_{\Upsilon,\mathfrak a}
    =
    \mathcal J^{(n)}_{\Upsilon}
    \,
    \frac{
    \left\langle
        \prod_{k=0}^{n-1}
        \Upsilon(\xi_{-,k},\bar\xi_{-,k})
        \Upsilon^\dagger(\xi_{+,k},\bar\xi_{+,k})
    \right\rangle_{\cD_{\mathfrak a}}
    }{
    \left[
    \left\langle
        \Upsilon(\eta_-,\bar\eta_-)
        \Upsilon^\dagger(\eta_+,\bar\eta_+)
    \right\rangle_{\cD_{\mathfrak a}}
    \right]^n
    } .
    \label{eq:F-general-disk}
\end{equation}
Here the Jacobian factor is
\begin{equation}
    \mathcal J^{(n)}_{\Upsilon}
    =
    \frac{
    \prod_{k=0}^{n-1}
    \left[
        (\xi'_{-,k})^h
        (\bar\xi'_{-,k})^{\bar h}
        (\xi'_{+,k})^h
        (\bar\xi'_{+,k})^{\bar h}
    \right]
    }{
    \left[
        (\eta'_{-})^h
        (\bar\eta'_{-})^{\bar h}
        (\eta'_{+})^h
        (\bar\eta'_{+})^{\bar h}
    \right]^n
    } .
    \label{eq:jacobian-general}
\end{equation}
Equation~\eqref{eq:F-general-disk} is the desired general CFT prediction for
the fixed-outcome post-measurement entanglement of a primary excited state.

For later use, we stress that the cancellation of the Jacobian factors from the first map
\(w\mapsto z\) is meant for a single primary field with fixed conformal weights
\((h,\bar h)\).  For a coherent superposition of fields with different spin, the common
overall factors still cancel in the normalized ratio, but the relative Jacobian between the
different chiral components may survive as a phase convention for the superposition
coefficients.  This point will be important for the \(J/\bar J\) superposition below.

Combining Eq.~\eqref{eq:F-general-disk} with
Eq.~\eqref{eq:S-excited-general}, the post-measurement R\'enyi entropy of the
excited state is
\begin{equation}
    S^{(\Upsilon,M)}_n(l,s,L)
    =
    S^{(0,M)}_n(l,s,L)
    +
    \frac{1}{1-n}
    \log F^{(n)}_{\Upsilon,\mathfrak a}(l,s,L).
    \label{eq:final-excited-renyi-section}
\end{equation}
This formula separates the universal geometry of the measured ground state
from the universal operator-dependent correction due to the excitation.
\section{The chiral current excitation in free compact boson CFT}
\label{sec:current-excitation}
\subsection{The free compact boson CFT}\label{sec:CFT-convention}
Although the geometric part of the construction in Sec.~\ref{sec:excited-replicated-geometry} is general, 
we now specialize to the compact free boson CFT and fix our conventions.  
The theory is defined by a real scalar field \(\varphi\) compactified as 
\(\varphi \sim \varphi+2\pi R\), with Euclidean action
\begin{equation}
    \mathcal{A}=\frac{1}{8\pi}\int d^2x\,(\partial_\mu \varphi)^2 .
\end{equation}
In complex coordinates the field decomposes into holomorphic and antiholomorphic parts,
\begin{equation}
    \varphi(z,\bar z)=\phi(z)+\bar\phi(\bar z),
\end{equation}
with the normalization
\begin{equation}
    \phi(z)\phi(w)\sim -\log(z-w),
    \qquad
    \bar\phi(\bar z)\bar\phi(\bar w)\sim -\log(\bar z-\bar w).
\end{equation}
The holomorphic and antiholomorphic \(U(1)\) currents are therefore
\begin{equation}
    J(z)=\ii\partial\phi(z),
    \qquad
    \bar J(\bar z)=\ii\bar\partial\bar\phi(\bar z),
\end{equation}
so that
\begin{equation}
    J(z)J(w)\sim \frac{1}{(z-w)^2},
    \qquad
    \bar J(\bar z)\bar J(\bar w)\sim \frac{1}{(\bar z-\bar w)^2}.
\end{equation}
The primary fields of the compact boson include the vertex operators
\begin{equation}
    V_{\alpha,\bar\alpha}(z,\bar z)
    =
    :\exp\!\left[\ii\alpha\phi(z)+\ii\bar\alpha\bar\phi(\bar z)\right]: ,
\end{equation}
where the allowed charges \((\alpha,\bar\alpha)\) are restricted by the compactification radius. With the above normalization, their conformal weights are
\begin{equation}
    h=\frac{\alpha^2}{2},
    \qquad
    \bar h=\frac{\bar\alpha^2}{2}.
\end{equation}
\subsection{The chiral current excitation}
We now specialize the general fixed-boundary formula to the holomorphic
current of the compact free boson,
\begin{equation}
    J(z)=\ii\partial\phi(z),\qquad (h,\bar h)=(1,0).
\end{equation}
This excitation is the CFT representative of the lowest right-moving
particle-hole excitation in the critical XX chain
\cite{AlcarazBerganzaSierra2011,BerganzaAlcarazSierra2012}. 

We assume that the boundary condition \(\mathfrak a\) preserves the
\(U(1)\) current algebra and that the current has no one-point function on
the disk. Then the current correlators on the uniformized disk are generated
by Wick contractions,
\begin{equation}
    \langle J(\xi_i)J(\xi_j)\rangle_{\mathbb D_{\mathfrak a}}
    =
    \frac{1}{(\xi_i-\xi_j)^2}.
    \label{eq:current-two-point}
\end{equation}
Thus the \(2n\)-point function is the hafnian
\begin{equation}
    \left\langle
        \prod_{i=1}^{2n}J(\xi_i)
    \right\rangle_{\mathbb D_{\mathfrak a}}
    =
    \Hf
    \left[
        \frac{1}{(\xi_i-\xi_{j})^2}
    \right]_{i,j=1}^{2n},
    \label{eq:current-hafnian}
\end{equation}
where the diagonal entries are set to zero, and
\begin{equation}
    \Hf(A)
    =
    \frac{1}{2^n n!}
    \sum_{\pi\in S_{2n}}
    \prod_{j=1}^{n}
    A_{\pi(2j-1),\pi(2j)} .
\end{equation}

The Jacobian factor is
\begin{equation}
    \frac{
        \prod_{k=0}^{n-1}\xi'_{-,k}\xi'_{+,k}
    }{
        (\eta'_-\eta'_+)^n
    }
    =
    n^{-2n}
    (\eta_-\eta_+)^{1-n}.
    \label{eq:current-jacobian}
\end{equation}
Dividing by the \(n\)-th power of the two-point function
\(\langle J(\eta_-)J(\eta_+)\rangle=(\eta_- - \eta_+)^{-2}\), we obtain
\begin{equation}
    F^{(n)}_{J,\mathfrak a}
    =
    n^{-2n}
    (\eta_-\eta_+)^{1-n}
    (\eta_- - \eta_+)^{2n}
    \Hf
    \left[
        \frac{1}{(\xi_i-\xi_{j})^2}
    \right]_{i,j=1}^{2n}.
    \label{eq:F-current-general}
\end{equation}
The \(2n\) points entering the hafnian are ordered as
\begin{equation}
    \{\xi_i\}_{i=1}^{2n}
    =
    \{
        \xi_{-,0},\xi_{+,0},
        \xi_{-,1},\xi_{+,1},
        \dots,
        \xi_{-,n-1},\xi_{+,n-1}
    \}.
\end{equation}
The post-measurement R\'enyi entropy of the current state is therefore
\begin{equation}
    S^{(J,M)}_n(l,s,L)
    =
    S^{(0,M)}_n(l,s,L)
    +
    \frac{1}{1-n}
    \log F^{(n)}_{J,\mathfrak a}(l,s,L).
    \label{eq:S-current-renyi}
\end{equation}
For the compact boson, \(c=1\), the ground-state piece is
\begin{equation}
    S^{(0,M)}_n(l,s,L)
    =
    \frac{1}{12}
    \left(
        1+\frac{1}{n}
    \right)
    \log
    \left[
        \frac{L}{\pi \epsilon}
        \frac{
            \sin\frac{\pi(l+s)}{L}
            \sin\frac{\pi l}{L}
        }{
            \sin\frac{\pi s}{L}
        }
    \right]
    +
    \gamma_{n,\mathfrak a}.
    \label{eq:current-ground-piece}
\end{equation}

\subsection{Second R\'enyi entropy}

For \(n=2\), set \(a=\eta_-^{1/2}\) and \(b=\eta_+^{1/2}\). The four
insertion points are \(\{\xi_i\}_{i=1}^4\)=\(\{a,b,-a,-b\}\), and Wick contractions or the hafnian give
\begin{equation}
    \langle J(a)J(b)J(-a)J(-b)\rangle
    =
    \frac{1}{(a-b)^4}
    +
    \frac{1}{(a+b)^4}
    +
    \frac{1}{16a^2b^2}.
    \label{eq:current-four-point}
\end{equation}
The Jacobian factor is
\begin{equation}
    \frac{
        \xi'_{-,0}\xi'_{+,0}\xi'_{-,1}\xi'_{+,1}
    }{
        (\eta'_-\eta'_+)^2
    }
    =
    \frac{1}{16a^2b^2}.
\end{equation}
Using \(\eta_- - \eta_+=a^2-b^2\), Eq.~\eqref{eq:F-current-general} gives
\begin{equation}
    F^{(2)}_{J,\mathfrak a}
    =
    \frac{(a^2-b^2)^4}{16a^2b^2}
    \left[
        \frac{1}{(a-b)^4}
        +
        \frac{1}{(a+b)^4}
        +
        \frac{1}{16a^2b^2}
    \right].
    \label{eq:F2-current-rq}
\end{equation}
Equivalently,
\begin{equation}
    F^{(2)}_{J,\mathfrak a}
    =
    \frac{(a+b)^4+(a-b)^4}{16a^2b^2}
    +
    \frac{(a^2-b^2)^4}{256a^4b^4}.
    \label{eq:F2-current-rq-simplified}
\end{equation}
Introducing the ratio \(\rho=\eta_+/\eta_-=b^2/a^2\), this simplifies to
\begin{equation}
    F^{(2)}_{J,\mathfrak a}
    =
    \frac{1}{256}
    \left(
        \rho+\rho^{-1}+14
    \right)^2 .
    \label{eq:F2-current-rho}
\end{equation}
Therefore
\begin{equation}
    S^{(J,M)}_2
    =
    S^{(0,M)}_2
    -
    \log F^{(2)}_{J,\mathfrak a}.
\end{equation}

\subsection{Third R\'enyi entropy}

For \(n=3\), let \(\omega=e^{2\pi \ii/3}\), \(r=\eta_-^{1/3}\), and
\(t=\eta_+^{1/3}\). The six insertion points are 
\(\{\xi_i\}_{i=1}^6=\{r,t,\omega r,\omega t,\omega^2r,\omega^2t\}\). The corresponding six-point
current correlator is
\begin{equation}
    \Hf_3(r,t)
    =
    \Hf
    \left[
        \frac{1}{(\xi_i-\xi_{j})^2}
    \right]_{i,j=1}^{6}.
\end{equation}
A direct evaluation gives
\begin{equation}
    \Hf_3(r,t)
    =
    \frac{
        9\left(r^6+7r^3t^3+t^6\right)^2
    }{
        (r-t)^6(r^2+rt+t^2)^6
    }.
    \label{eq:Hf3-result}
\end{equation}
Then substituting this into Eq.~\eqref{eq:F-current-general} with \(n=3\), we get
\begin{equation}
    F^{(3)}_{J,\mathfrak a}
    =
    3^{-6}
    (\eta_-\eta_+)^{-2}
    (\eta_- - \eta_+)^6
    \Hf_3(r,t).
\end{equation}
Since \(\eta_- - \eta_+=r^3-t^3=(r-t)(r^2+rt+t^2)\), the denominator in
Eq.~\eqref{eq:Hf3-result} cancels against the factor
\((\eta_- - \eta_+)^6\) in the equation above, and we finally obtain
\begin{equation}
    F^{(3)}_{J,\mathfrak a}
    =
    \frac{1}{81}
    \left(
        \frac{r^3}{t^3}
        +7+
        \frac{t^3}{r^3}
    \right)^2.
\end{equation}
Using \(\rho=\eta_+/\eta_-=t^3/r^3\), we obtain
\begin{equation}
    F^{(3)}_{J,\mathfrak a}
    =
    \frac{1}{81}
    \left(
        \rho+\rho^{-1}+7
    \right)^2 .
    \label{eq:F3-current-rho}
\end{equation}
For real \(l,s,L\), \(|\rho|=1\). Writing \(\rho=e^{2\ii\Theta}\), this may be
written as
\begin{equation}
    F^{(3)}_{J,\mathfrak a}
    =
    \frac{1}{81}
    \left(
        7+2\cos 2\Theta
    \right)^2 .
\end{equation}
The third R\'enyi entropy is
\begin{equation}
    S^{(J,M)}_3
    =
    S^{(0,M)}_3
    -
    \frac12
    \log F^{(3)}_{J,\mathfrak a}.
\end{equation}

\subsection{Recovery of the unmeasured-cylinder result}
\label{subsec:current-unmeasured-limit}

The slit-cylinder geometry should not be reduced to the ordinary cylinder by
setting the slit length to zero at fixed cutoff. The two geometries have
different topology. The correct prescription is to set the slit length equal
to the UV cutoff, \(s=\epsilon\), and then take \(\epsilon/L\to0\) with
\(x=l/L\) fixed. The UV cutoff does not appear explicitly in
\(F^{(n)}_{J,\mathfrak a}\), because this ratio is normalized by the
ground-state contribution; it enters only through
\(\alpha_s=\pi s/(2L)=\pi \epsilon/(2L)\).

Then 
\begin{equation}
    y=
    \left[
        \frac{\sin(\pi x+2\alpha_s)}{\sin\pi x}
    \right]^{1/2}.
\end{equation}
Recall that
\begin{equation}
    \eta_-=
    \frac{e^{\ii\alpha_s}-y}{e^{\ii\alpha_s}+y},
    \qquad
    \eta_+=
    \frac{e^{-\ii\alpha_s}-y}{e^{-\ii\alpha_s}+y},
\end{equation}
as \(\alpha_s\to0^+\), one has
\begin{equation}
    y=1+\alpha_s\cot\pi x+O(\alpha_s^2),
    \qquad
    \eta_-=\frac{\alpha_s(\ii-\cot\pi x)}{2}+O(\alpha_s^2),
    \qquad
    \eta_+=\frac{\alpha_s(-\ii-\cot\pi x)}{2}+O(\alpha_s^2).
\end{equation}
Although both \(\eta_-\) and \(\eta_+\) vanish as \(\alpha_s\to0^+\), their ratio remains finite:
\begin{equation}
    \rho
    =
    \frac{\eta_+}{\eta_-}
    =
    \frac{-\ii-\cot\pi x}{\ii-\cot\pi x}
    +O(\alpha_s)
    =
    e^{2\pi\ii x}
    +O(\alpha_s).
    \label{eq:rho-unmeasured-limit}
\end{equation}
Hence, in the continuum limit with \(s=\epsilon\), we have \(\rho\rightarrow e^{2\pi \ii x}\).
Substituting this into Eq.~\eqref{eq:F2-current-rho} gives
\begin{equation}
    F^{(2)}_{J}
    =
    \frac{1}{256}
    \left(
        14+2\cos 2\pi x
    \right)^2
    =
    \frac{1}{64}
    \left(
        7+\cos 2\pi x
    \right)^2 .
    \label{eq:F2-current-unmeasured}
\end{equation}
Equivalently, with \(r=\sin(\pi x/2)\),
\begin{equation}
    F^{(2)}_{J}
    =
    1-2r^2+3r^4-2r^6+r^8 .
    \label{eq:F2-current-polynomial}
\end{equation}
This is the standard \(n=2\) current result on the ordinary cylinder
\cite{BerganzaAlcarazSierra2012}. Similarly,
\begin{equation}
    F^{(3)}_{J}
    =
    \frac{1}{81}
    \left(
        7+2\cos 2\pi x
    \right)^2 ,
    \label{eq:F3-current-unmeasured}
\end{equation}
in agreement with the normalized current result obtained from the
ordinary-cylinder determinant or hafnian formula.

\section{Superposition of \(J\) and \(\bar J\)}
\label{sec:current-superposition}

In this section, we will consider the following current superposition state defined on the cylinder
\begin{equation}
|\Psi_1\rangle
=
\sqrt{p}\,|J\rangle
+
e^{\ii\theta_{\rm cyl}}\sqrt{q}\,|\bar J\rangle,
\qquad
q\equiv 1-p,
\qquad
J=\ii\partial\phi,\quad
\bar J=\ii\bar\partial\bar\phi .
\label{eq:JJbar-state}
\end{equation}

Before applying the general construction of Sec.~\ref{sec:excited-replicated-geometry}, we should slightly refine the
meaning of the phase in Eq.~(\ref{eq:JJbar-state}).  The formula (\ref{eq:F-general-disk}) was derived for an excitation
created by a single primary field with fixed conformal weights \((h,\bar h)\).  In that
case, under the first conformal map from the slit cylinder to the upper half-plane, the
external Jacobian factors in the numerator are exactly cancelled by those in the
denominator raised to the \(n\)-th power.  For the present state this cancellation cannot
be applied directly, because the two components of the superposition have different
chiral weights,
\begin{equation}
J:\ (h,\bar h)=(1,0),
\qquad
\bar J:\ (h,\bar h)=(0,1).
\end{equation}
Thus the common overall Jacobian still cancels, but the relative Jacobian between the
holomorphic and antiholomorphic components has to be kept.

Under the first map \(z=z(w)\), the two currents transform as
\begin{equation}
J_w(w_\mu)=z'(w_\mu)J_z(z_\mu),
\qquad
\bar J_w(\bar w_\mu)=\bar z'(\bar w_\mu)\bar J_z(\bar z_\mu),
\qquad
\mu=\pm .
\end{equation}
Therefore the relative factors between the \(\bar J\) and \(J\) components are
\begin{equation}
r_\mu(s)=
\frac{\bar z'(\bar w_\mu)}{z'(w_\mu)} .
\end{equation}
For the slit-cylinder map one finds, near \(w_\pm=\pm \ii\infty\),
\begin{equation}
r_-(s)=-e^{-2\ii\alpha_s},
\qquad
r_+(s)=-e^{2\ii\alpha_s},
\end{equation}
where \(\alpha_s=\frac{\pi s}{2L}\). The common minus sign is already present in the reference geometry \(s=0\):
\begin{equation}
r_-(0)=r_+(0)=-1.
\end{equation}
It is therefore a phase associated with the choice of local coordinates at the
infinite-cylinder insertions, rather than a measurement-induced phase.  The
geometry-dependent relative phase is obtained by comparing with this reference
convention,
\begin{equation}
\frac{r_-(s)}{r_-(0)}=e^{-2\ii\alpha_s},
\qquad
\frac{r_+(s)}{r_+(0)}=e^{2\ii\alpha_s}.
\end{equation}
Consequently, if \(\theta_{\rm cyl}\) denotes the relative phase in the cylinder
state with the same \(s=0\) convention, the phase entering the disk correlators is
\begin{equation}
\theta
=
\theta_{\rm cyl}-2\alpha_s
=
\theta_{\rm cyl}-\frac{\pi s}{L}.
\end{equation}

\subsection{The general formula}
We use the geometric variables \(\alpha_s,x,y,\eta_\pm,\xi_{\pm,k}\) 
defined in Sec. \ref{sec:slit-geometry}. The new ingredient in the \(J/\bar J\) superposition 
is the antiholomorphic image under the boundary doubling map:
\begin{equation}
\bar\eta(\bar z)=\frac{\bar z+\ii y}{\bar z-\ii y},
\qquad
\bar\eta_\pm=\bar\eta(\bar z_\pm),
\qquad
\eta'_\pm=\frac{2\ii y}{(z_\pm+\ii y)^2},
\qquad
\bar\eta'_\pm=-\frac{2\ii y}{(\bar z_\pm-\ii y)^2}.
\label{eq:Jbar-eta-derivatives}
\end{equation}
For real \(l,s,L\), \(\bar\eta_\pm=\overline{\eta_\pm}\). The mirror point of a disk coordinate \(\xi\) is \(\xi^\star=1/\bar\xi\). The boundary condition \(\mathfrak{a}\) glues the antiholomorphic current to the image holomorphic current as
\begin{equation}
\bar J(\bar\xi)\longrightarrow
\lambda_\mathfrak{a}\left(\frac{d\xi^\star}{d\bar\xi}\right)J(\xi^\star),
\qquad
\lambda_\mathfrak{a}=\pm1 .
\label{eq:Jbar-gluing}
\end{equation}
The sign \(\lambda_\mathfrak{a}\) is fixed by the conformal boundary condition selected by the measurement outcome.

For the insertion labeled by \((\mu,k)\), with \(\mu=-,+\), the holomorphic disk position \(\xi_{\mu,k}\), mirror position \(\xi^\star_{\mu,k}\), and the Jacobians \(A_{\mu,k}, \bar A_{\mu,k}\) are given by
\begin{equation}\label{eq:xixistar}
\begin{split}
&\xi_{\mu,k}=e^{2\pi \ii k/n}\eta_\mu^{1/n},
\qquad
\xi^\star_{\mu,k}=e^{2\pi \ii k/n}\bar\eta_\mu^{-1/n},
\\
&A_{\mu,k}=\frac{1}{n}e^{2\pi \ii k/n}\eta_\mu^{1/n-1}\eta'_\mu,
\qquad
\bar A_{\mu,k}
=
-\frac{\lambda_\mathfrak{a}}{n}e^{2\pi \ii k/n}\bar\eta_\mu^{-1/n-1}\bar\eta'_\mu .
\end{split}
\end{equation}
The coefficients multiplying \(J\) and \(\bar J\) are
\begin{equation}
c^J_-=\sqrt p,\qquad c^J_+=\sqrt p,\qquad
c^{\bar J}_-=e^{\ii\theta}\sqrt{q},\qquad
c^{\bar J}_+=e^{-\ii\theta}\sqrt{q} .
\label{eq:Jbar-general-coefficients}
\end{equation}

After doubling, each insertion of \(\Psi\) is a linear combination of two holomorphic currents, one at \(\xi_{\mu,k}\) and one at \(\xi^\star_{\mu,k}\). Since the current theory is Gaussian, the replicated \(2n\)-point function is a hafnian. For two insertions \(I=(\mu,k)\) and \(J=(\nu,l)\), define
\begin{equation}
\begin{aligned}
{\cal G}_{IJ}={}&
c^J_\mu c^J_\nu
\frac{A_{\mu,k}A_{\nu,l}}{(\xi_{\mu,k}-\xi_{\nu,l})^2}
+
c^J_\mu c^{\bar J}_\nu
\frac{A_{\mu,k}\bar A_{\nu,l}}{(\xi_{\mu,k}-\xi^\star_{\nu,l})^2}
\\
&+
c^{\bar J}_\mu c^J_\nu
\frac{\bar A_{\mu,k}A_{\nu,l}}{(\xi^\star_{\mu,k}-\xi_{\nu,l})^2}
+
c^{\bar J}_\mu c^{\bar J}_\nu
\frac{\bar A_{\mu,k}\bar A_{\nu,l}}{(\xi^\star_{\mu,k}-\xi^\star_{\nu,l})^2}.
\end{aligned}
\label{eq:Jbar-general-kernel}
\end{equation}
The diagonal entries are set to zero. With the ordering
\begin{equation}
\{I\}=\{(-,0),(+,0),(-,1),(+,1),\ldots,(-,n-1),(+,n-1)\},
\label{eq:Jbar-Hafnian-ordering}
\end{equation}
the \(2n\)-point correlator in the numerator of Eq.~(\ref{eq:F-general-disk}) is given by \(\operatorname{Hf}{\cal G}\).

The single-sheet correlator in the denominator of Eq.~(\ref{eq:F-general-disk}) is obtained using the same expression in Eq.~(\ref{eq:xixistar}) evaluated at \(n=1\):
\begin{equation}
\eta_\pm^\star=\frac{1}{\bar\eta_\pm},
\qquad
A_\pm^{(1)}=\eta'_\pm,
\qquad
\bar A_\pm^{(1)}=-\lambda_\mathfrak{a}\frac{\bar\eta'_\pm}{\bar\eta_\pm^2}.
\label{eq:Jbar-single-sheet-data}
\end{equation}
Then
\begin{equation}
\begin{aligned}
{\cal G}^{(1)}_{-+}={}&
p\,\frac{A_-^{(1)}A_+^{(1)}}{(\eta_- - \eta_+)^2}
+\sqrt{pq}\,e^{-\ii\theta}
\frac{A_-^{(1)}\bar A_+^{(1)}}{(\eta_- - \eta_+^\star)^2}
\\
&+\sqrt{pq}\,e^{\ii\theta}
\frac{\bar A_-^{(1)}A_+^{(1)}}{(\eta_-^\star-\eta_+)^2}
+q\,\frac{\bar A_-^{(1)}\bar A_+^{(1)}}{(\eta_-^\star-\eta_+^\star)^2}.
\end{aligned}
\label{eq:Jbar-single-sheet-kernel}
\end{equation}
The general \(n\)-replica post-measurement ratio for the current superposition state is therefore
\begin{equation}
F^{(n)}_{\Psi_1,\mathfrak{a}}(l,s,L;p,\theta,\lambda_{\mathfrak{a}})
=
\frac{\operatorname{Hf}{\cal G}}{\left({\cal G}^{(1)}_{-+}\right)^n}.
\label{eq:Jbar-general-Fn}
\end{equation}
This formula already includes the conformal Jacobians, since they are contained in the dressed coefficients \(A_{\mu,k}\) and \(\bar A_{\mu,k}\).

\subsection{\texorpdfstring{\(n=2\) formula}{n=2 formula}}
\label{subsec:Jbar-n2}

For \(n=2\), set \(a=\eta_-^{1/2},b=\eta_+^{1/2}\) as before, then the four insertion points are given by \(\{\xi_i\}_{i=1}^4\)=\(\{a,b,-a,-b\}\).
The points \(\xi_1,\xi_3\) are incoming insertions and \(\xi_2,\xi_4\) are outgoing insertions. Introduce
\begin{equation}
\varepsilon_1=\varepsilon_3=+1,
\qquad
\varepsilon_2=\varepsilon_4=-1,
\qquad
s_1=s_2=+1,
\qquad
s_3=s_4=-1,
\label{eq:Jbar-n2-signs}
\end{equation}
where \(\varepsilon_i\) tracks the phase of the \(\bar J\) component and \(s_i\) tracks the sheet sign. Let \(\mu_i=-\) for \(i=1,3\) and \(\mu_i=+\) for \(i=2,4\). Using the order in Eq.~\eqref{eq:Jbar-Hafnian-ordering}, the variables in Eq.~\eqref{eq:xixistar} become
\begin{equation}
\xi_i=s_i\eta_{\mu_i}^{1/2},
\qquad
\xi_i^\star=s_i\bar\eta_{\mu_i}^{-1/2},
\qquad
A_i=s_i\frac{\eta'_{\mu_i}}{2\eta_{\mu_i}^{1/2}},
\qquad
\bar A_i=-\lambda_{\mathfrak{a}} s_i\frac{\bar\eta'_{\mu_i}}{2\bar\eta_{\mu_i}^{3/2}} .
\label{eq:Jbar-n2-data}
\end{equation}
With the notation of \eqref{eq:Jbar-n2-signs} and \eqref{eq:Jbar-n2-data}, the \(n=2\) doubled-current kernel is
\begin{equation}
\begin{aligned}
{\cal G}_{ij}={}&
p\,\frac{A_iA_j}{(\xi_i-\xi_j)^2}
+\sqrt{pq}\,e^{\ii\varepsilon_j\theta}
\frac{A_i\bar A_j}{(\xi_i-\xi_j^\star)^2}\\
&+\sqrt{pq}\,e^{\ii\varepsilon_i\theta}
\frac{\bar A_iA_j}{(\xi_i^\star-\xi_j)^2}
+q\,e^{\ii(\varepsilon_i+\varepsilon_j)\theta}
\frac{\bar A_i\bar A_j}{(\xi_i^\star-\xi_j^\star)^2}.
\end{aligned}
\label{eq:Jbar-n2-kernel}
\end{equation}
The four-point function is the hafnian of this \(4\times4\) kernel,
\begin{equation}
{\cal Z}^{J\bar J}_2
=
{\cal G}_{12}{\cal G}_{34}
+
{\cal G}_{13}{\cal G}_{24}
+
{\cal G}_{14}{\cal G}_{23}.
\label{eq:Jbar-n2-Z2}
\end{equation}
The single-sheet denominator \({\cal Z}^{J\bar J}_1={\cal G}^{(1)}_{-+}\) and the explicit expression was given in \eqref{eq:Jbar-single-sheet-kernel}.
Using the explicit single-sheet coordinates, the four terms in Eq.~\eqref{eq:Jbar-single-sheet-kernel}
simplify to
\begin{equation}
{\cal Z}_1^{J\bar J}
=
\frac{1}{4\sin^2\alpha_s}
-
\frac{\lambda_{\mathfrak{a}}}{2}\sqrt{pq}\cos(\theta+2\alpha_s).
\end{equation}
Thus the \(n=2\) ratio is
\begin{equation}
F^{(2)}_{\Psi_1,\mathfrak{a}}(l,s,L;p,\theta,\lambda_{\mathfrak{a}})
=
\frac{
{\cal G}_{12}{\cal G}_{34}
+
{\cal G}_{13}{\cal G}_{24}
+
{\cal G}_{14}{\cal G}_{23}
}{
\left({\cal Z}^{J\bar J}_1\right)^2
}.
\label{eq:Jbar-n2-F2}
\end{equation}
The corresponding second R\'enyi entropy shift is
\begin{equation}
S^{(\Psi_1,M)}_2(l,s,L;p,\theta)-S^{(0,M)}_2(l,s,L)
=
-\log F^{(2)}_{\Psi_1,\mathfrak{a}}(l,s,L;p,\theta,\lambda_{\mathfrak{a}}).
\label{eq:Jbar-entropy-shift}
\end{equation}

Several checks are immediate. For \(p=1\), the state reduces to the holomorphic current and \eqref{eq:Jbar-n2-F2} gives
\begin{equation}
F^{(2)}_{J}
=
\frac{1}{256}\left(\rho+\rho^{-1}+14\right)^2,
\qquad
\rho=\frac{\eta_+}{\eta_-}.
\label{eq:Jbar-pure-J-check}
\end{equation}
For \(p=0\), the state reduces to the antiholomorphic current and gives \(F^{(2)}_{\bar J}=F^{(2)}_{J}\). For \(0<p<1\), the mixed terms in \eqref{eq:Jbar-n2-kernel} are generically nonzero and the ratio depends on the relative phase \(\theta\).

\section{Conjugate-vertex superposition states}
\label{sec:vertex-superposition}
In this section we will consider the superposition state of a vertex operator and its conjugate, 
\begin{equation}
V_{\boldsymbol\alpha}(\xi,\bar\xi)
\equiv
V_{\alpha,\bar\alpha}(\xi,\bar\xi)
=
:e^{\ii\alpha\phi(\xi)+\ii\bar\alpha\bar\phi(\bar\xi)}:,
\qquad
V_{-\boldsymbol\alpha}=V_{\boldsymbol\alpha}^{\dagger}.
\label{eq:general-vertex-def}
\end{equation}
Here we keep the notation parallel to the current-superposition state in the previous section.
It will be useful to denote \(\mathcal{V}_{\sigma}\equiv V_{\sigma\alpha,\sigma\bar\alpha}\), with \(\sigma=\pm1\). The excited operator is the coherent superposition \footnote{Similar superposition states were studied in Ref.~\cite{Chen:2023gql} in the context of entanglement asymmetry.}
\begin{equation}
\Psi_2
=
\sqrt p\,\mathcal{V}_{+}
+
e^{\ii\theta}\sqrt{q}\,\mathcal{V}_{-},
\qquad
\Psi_2^\dagger
=
\sqrt p\,\mathcal{V}_{-}
+
e^{-\ii\theta}\sqrt{q}\,\mathcal{V}_{+} .
\label{eq:general-vertex-superposition}
\end{equation}
We note that the phase convention issue discussed for the \(J/\bar J\) superposition
does not arise for the present vertex superposition.  Although \(\mathcal{V}_{+}\) and \(\mathcal{V}_{-}\)
carry opposite charges, they have the same conformal weights,
\begin{equation}
h_{+}=h_{-}=\frac{\alpha^2}{2},
\qquad
\bar h_{+}=\bar h_{-}=\frac{\bar\alpha^2}{2}.
\end{equation}
Therefore the Jacobian factors produced by the first map \(w\mapsto z\) are identical
for the two components of the superposition.  They cancel in the normalized replica
ratio in the same way as for a single primary field.  Consequently, no additional
relative phase has to be introduced before evaluating the disk correlation functions.
\subsection{The general formula}
The boundary condition on the disk is implemented by the gluing rule
\begin{equation}
\bar\phi(\bar\xi)\longrightarrow \lambda_{\mathfrak{a}}\,\phi(\xi^\star)+\chi_{\mathfrak{a}},
\qquad
\xi^\star=\frac{1}{\bar\xi},
\qquad
\lambda_{\mathfrak{a}}=\pm1 .
\label{eq:general-vertex-gluing}
\end{equation}

The constant $\chi_{\mathfrak{a}}$ denotes the zero-mode parameter of the compact-boson
boundary condition. After the doubling trick, we eliminate the image coordinates using
\(\xi_i^\star=1/\bar\xi_i\). Contractions between an insertion and the image
of another insertion then generate the disk factors
\(1-\xi_i\bar\xi_j\) and \(1-\xi_j\bar\xi_i\), while the self-image
contractions generate the one-body factors \(1-|\xi_i|^2\). With the usual
choice of vertex-operator cocycles and branches, the disk correlator can
therefore be written as \footnote{For example,
\(\xi_i-\xi_j^\star=-(1-\xi_i\bar\xi_j)/\bar\xi_j\) and
\(\xi_i^\star-\xi_j^\star=-(\bar\xi_i-\bar\xi_j)/(\bar\xi_i\bar\xi_j)\).
The extra powers of \(\bar\xi_i\) cancel after including the inversion
Jacobian and the neutrality condition.}
\begin{equation}
\left\langle \prod_{i=1}^m \mathcal{V}_{\sigma_i}(\xi_i,\bar\xi_i)\right\rangle_{\mathbb{D}_{\mathfrak{a}}}
=
{\cal S}_{\mathfrak{a}}(\boldsymbol\sigma)
\prod_{i=1}^m(1-|\xi_i|^2)^{\lambda_{\mathfrak{a}}\alpha\bar\alpha}
\prod_{1\le i<j\le m}
{\cal K}_{\alpha,\bar\alpha}^{(\lambda_{\mathfrak{a}})}(\xi_i,\xi_j)^{\sigma_i\sigma_j},
\label{eq:general-disk-vertex-correlator}
\end{equation}
where
\begin{equation}
{\cal K}_{\alpha,\bar\alpha}^{(\lambda_{\mathfrak{a}})}(x,y)
=
(x-y)^{\alpha^2}(\bar x-\bar y)^{\bar \alpha^2}(1-x\bar y)^{\lambda_{\mathfrak{a}}\alpha\bar\alpha}(1-y\bar x)^{\lambda_{\mathfrak{a}}\alpha\bar\alpha} .
\label{eq:general-disk-kernel}
\end{equation}
The factor \({\cal S}_{\mathfrak{a}}(\boldsymbol\sigma)\) is the zero-mode selection rule. In the doubled chiral description it can be written as
\begin{equation}
{\cal S}_{\mathfrak{a}}(\boldsymbol\sigma)
=\exp\!\left(\ii\bar\alpha\chi_{\mathfrak{a}}\sum_i\sigma_i\right)
\delta_{(\alpha+\lambda_{\mathfrak{a}}\bar\alpha)\sum_i\sigma_i,\,0}.
\label{eq:general-selection-rule}
\end{equation}
When $\alpha+\lambda_{\mathfrak{a}}\bar\alpha\neq0$, the neutrality condition implies $\sum_i\sigma_i=0$, so the $\chi_{\mathfrak{a}}$-dependent phase drops out.

We use the same slit-cylinder geometry and the same variables as in Sec.~\ref{sec:current-superposition}. The variables \(\eta_\pm\) are given by Eq.~\eqref{eq:eta-pm}. For the \(n\)-sheeted surface, the \(2n\) insertion points on the disk are
\begin{equation}
\xi_{-,k}=e^{2\pi \ii k/n}\eta_-^{1/n},
\qquad
\xi_{+,k}=e^{2\pi \ii k/n}\eta_+^{1/n},
\qquad
k=0,\ldots,n-1 .
\label{eq:general-n-insertion-points}
\end{equation}
We assign the coefficients
\begin{equation}
c_-(+)=\sqrt p,\qquad c_-(-)=e^{\ii\theta}\sqrt{q},
\qquad
c_+(+)=e^{-\ii\theta}\sqrt{q},\qquad c_+(-)=\sqrt p .
\label{eq:general-coefficients}
\end{equation}
With these coefficients, the numerator of the normalized replica
correlator is obtained by expanding the superposed operator on every
replica. Before separating off the Jacobian and self-image factors, its
charge-dependent part has the structure
\begin{equation}
\left\langle
\prod_{k=0}^{n-1}
\left[
\sum_{\sigma_{-,k}=\pm}
c_-(\sigma_{-,k})\mathcal{V}_{\sigma_{-,k}}(\xi_{-,k},\bar\xi_{-,k})
\right]
\left[
\sum_{\sigma_{+,k}=\pm}
c_+(\sigma_{+,k})\mathcal{V}_{\sigma_{+,k}}(\xi_{+,k},\bar\xi_{+,k})
\right]
\right\rangle_{\cD_{\mathfrak a}}.
\end{equation}
Here \(\sigma_{-,k}\) labels the charge chosen from the operator creating
the ket on the \(k\)-th sheet, while \(\sigma_{+,k}\) labels the charge
chosen from the conjugate operator in the bra. Expanding the products gives
a sum over \(2^{2n}\) charge sectors. For a fixed sector
\begin{equation}
\boldsymbol\sigma=\{\sigma_{-,0},\sigma_{+,0},\ldots,
\sigma_{-,n-1},\sigma_{+,n-1}\},
\end{equation}
the disk correlator in Eq.~\eqref{eq:general-disk-vertex-correlator} contributes the zero-mode factor
\(S_a(\sigma)\) and the pairwise product of kernels between all distinct
insertions. Hence the charge-sector sum entering the \(n\)-replica
numerator is
\begin{equation}
{\cal Z}_n
=
\sum_{\boldsymbol\sigma}
{\cal S}_\mathfrak{a}(\boldsymbol\sigma)
\prod_{k=0}^{n-1}
c_-(\sigma_{-,k})c_+(\sigma_{+,k})
\prod_{(r,\mu)<(s,\nu)}
{\cal K}_{\alpha,\bar\alpha}^{(\lambda_{\mathfrak{a}})}
(\xi_{\mu,r},\xi_{\nu,s})^{\sigma_{\mu,r}\sigma_{\nu,s}},
\label{eq:general-Zn}
\end{equation}
where \(\mu,\nu=\pm\), and the last product runs over all unordered pairs of distinct insertions. The ordering condition \((r,\mu)<(s,\nu)\) means that each unordered pair of distinct insertions is included once. The corresponding single-sheet quantity is obtained by setting \(n=1\). It is the same charge expansion with the two insertions located at
\(\eta_-\) and \(\eta_+\):
\begin{equation}
{\cal Z}_1
=
\sum_{\sigma,\tau=\pm1}
{\cal S}_\mathfrak{a}(\sigma,\tau)c_-(\sigma)c_+(\tau)
{\cal K}_{\alpha,\bar\alpha}^{(\lambda_{\mathfrak{a}})}(\eta_-,\eta_+)^{\sigma\tau}.
\label{eq:general-Z1}
\end{equation}
The charge sums \({\cal Z}_n\) and \({\cal Z}_1\) contain the pairwise contractions between different vertex insertions. Two charge-independent factors remain to be included. The first one is the Jacobian factor from the uniformizing map
\(\eta\mapsto \xi=e^{2\pi \ii k/n}\eta^{1/n}\). Since
\(V_{\alpha,\bar\alpha}\) has conformal weights
\begin{equation}
h=\frac{\alpha^2}{2},\qquad 
\bar h=\frac{\bar\alpha^2}{2},
\end{equation}
each insertion contributes a primary-field factor \(\left(\frac{d\xi_{\mu,k}}{d\eta_\mu}\right)^h
\left(\frac{d\bar\xi_{\mu,k}}{d\bar\eta_\mu}\right)^{\bar h}\), with
\begin{equation}
\frac{d\xi_{\mu,k}}{d\eta_\mu}
=
\frac{1}{n}e^{2\pi \ii k/n}\eta_\mu^{1/n-1},
\qquad
\mu=\pm .
\end{equation}
Multiplying over \(k=0,\ldots,n-1\) and over the two insertions \(\mu=\pm\), and absorbing the root-of-unity phases into the branch convention
for the vertex operators, gives
\begin{equation}
{\cal J}_n
=
n^{-n(\alpha^2+\bar\alpha^2)}
(\eta_-\eta_+)^{(1-n)\alpha^2/2}
(\bar\eta_-\bar\eta_+)^{(1-n)\bar\alpha^2/2}.
\end{equation}

The second factor comes from the self-image contractions in the disk correlator. On the \(n\)-sheeted surface,
\begin{equation}
|\xi_{\mu,k}|^2=|\eta_\mu|^{2/n},
\qquad \mu=\pm ,
\end{equation}
so the self-image part of the numerator is
\begin{equation}
\left[
(1-|\eta_-|^{2/n})(1-|\eta_+|^{2/n})
\right]^{n\lambda_{\mathfrak{a}}\alpha\bar\alpha}.
\end{equation}
The denominator contains the \(n\)-th power of the corresponding single-sheet factor,
\begin{equation}
\left[
(1-|\eta_-|^2)(1-|\eta_+|^2)
\right]^{n\lambda_{\mathfrak{a}}\alpha\bar\alpha}.
\end{equation}
Thus the normalized self-image contribution is
\begin{equation}
{\cal R}_n
=
\left[
\frac{
(1-|\eta_-|^{2/n})(1-|\eta_+|^{2/n})
}{
(1-|\eta_-|^2)(1-|\eta_+|^2)
}
\right]^{n\lambda_{\mathfrak{a}}\alpha\bar\alpha}.
\end{equation}
Combining the Jacobian factor, the self-image factor, and the charge sums, we obtain
\begin{equation}
F^{(n)}_{\Psi_2,\mathfrak{a}}(l,s,L;p,\theta,\lambda_{\mathfrak{a}})
=
{\cal J}_n {\cal R}_n \frac{{\cal Z}_n}{{\cal Z}_1^n}.
\label{eq:general-Fn-vertex}
\end{equation}
This formula is the most direct way to evaluate the post-measurement ratio for arbitrary \(n\). For generic boundary gluing and generic charges, the selection rule \eqref{eq:general-selection-rule} removes part of the charge sum. In the special case \(\alpha+\lambda_{\mathfrak{a}}\bar\alpha=0\), each bulk vertex is neutral after doubling and all \(2^{2n}\) sectors contribute.

\subsection{\(n=2\) formula for superposition of \(V_{1,-1}\) and \(V_{-1,1}\)}
\label{subsec:v11-n2}

We now specialize to the vertex pair realized by the two Umklapp excitations of the XX chain,
\begin{equation}
V_+=V_{1,-1},\qquad V_-=V_{-1,1}.
\label{eq:v1qair}
\end{equation}
Among the fixed measurement outcomes considered in the XX chain, the
antiferromagnetic state flows to a conformal boundary condition in the infrared.  In the compact-boson
convention of Sec.~\ref{sec:CFT-convention}, this boundary condition corresponds to a Neumann
condition for \(\varphi=\phi+\bar\phi\).
Equivalently, it is a Dirichlet condition for the dual field \(\widetilde{\varphi}=\phi-\bar\phi\),
and is implemented by the gluing rule
\begin{equation}
    \bar\phi(\bar\xi)\rightarrow \lambda_{\mathfrak{a}}\phi(\xi^\star)+\chi_{\mathfrak{a}},
    \qquad \lambda_{\mathfrak{a}}=+1 .
\end{equation}
Therefore
\begin{equation}
\alpha=1,\qquad \bar\alpha=-1,\qquad \lambda_{\mathfrak{a}}\alpha\bar\alpha=-1,\qquad \alpha+\lambda_{\mathfrak{a}}\bar\alpha=0 .
\label{eq:ABC-lambdaplus}
\end{equation}
For the present choice $\alpha=1$, $\bar\alpha=-1$, and $\lambda_{\mathfrak{a}}=+1$, so that $\alpha+\lambda_a\bar\alpha=0$ and the zero-mode phase is not removed by the neutrality condition. Instead, the two components of the superposition acquire opposite phases under the doubling map,
\begin{equation}
V_+ \longrightarrow e^{-\ii\chi_{\mathfrak{a}}}\widetilde V_+,
\qquad
V_- \longrightarrow e^{+\ii\chi_{\mathfrak{a}}}\widetilde V_-,
\end{equation}
where $\widetilde V_\pm$ denote the corresponding doubled operators with the constant zero-mode factor removed. Therefore
\begin{equation}
\sqrt{p}\,V_+
+e^{\ii\theta}\sqrt{q}\,V_-
\longrightarrow
e^{-\ii\chi_{\mathfrak{a}}}
\left[
\sqrt{p}\,\widetilde V_+
+
e^{\ii(\theta+2\chi_{\mathfrak{a}})}
\sqrt{q}\,\widetilde V_-
\right].
\end{equation}
The overall phase is irrelevant for the normalized replica ratio. Hence the boundary zero mode parameter enters only through the effective relative phase
\begin{equation}
\theta_{\rm eff}=\theta+2\chi_{\mathfrak{a}}.
\end{equation}
In the following we absorb this constant shift into the definition of the relative phase and denote $\theta_{\rm eff}$ simply by $\theta$.

All \(16\) charge sectors contribute. The disk kernel becomes
\begin{equation}
K(x,y)
\equiv
{\cal K}_{1,-1}^{(+1)}(x,y)
=
\frac{|x-y|^2}{|1-x\bar y|^2}.
\label{eq:kernel-lambdaplus}
\end{equation}

For \(n=2\), set \(a=\eta_-^{1/2}, b=\eta_+^{1/2}\), then
\begin{equation}
\xi_1=a,\quad \xi_2=b,\quad \xi_3=-a,\quad \xi_4=-b .
\label{eq:n2-points}
\end{equation}
The points \(\xi_1,\xi_3\) are incoming insertions and \(\xi_2,\xi_4\) are outgoing insertions. Since \(\eta_+=\overline{\eta_-}\) for real \(l,s,L\), the square-root branch can be chosen so that \(|a|=|b|\). The independent pairwise kernels are 
\begin{equation}
K_- = K(a,b)=\frac{|a-b|^2}{|1-a\bar b|^2},
\qquad
K_+ = K(a,-b)=\frac{|a+b|^2}{|1+a\bar b|^2},
\qquad
H = K(a,-a)=\frac{4|a|^2}{(1+|a|^2)^2}.
\label{eq:basic-kernels}
\end{equation}
The single-sheet kernel is
\begin{equation}
K_0=K(\eta_-,\eta_+)=K(a^2,b^2)=K_-K_+ .
\label{eq:v11-K0}
\end{equation}

The \(n=2\) Jacobian is \({\cal J}_2=1/(16|ab|^2)\). The self-image ratio is \({\cal R}_2=(1+|a|^2)^2(1+|b|^2)^2\). Hence
\begin{equation}
{\cal J}_2{\cal R}_2
=
\frac{(1+|a|^2)^2(1+|b|^2)^2}{16|ab|^2}
=
\frac{1}{H^2},
\label{eq:v1qrefactor}
\end{equation}
where we have used the fact that \(|a|=|b|\). 
We now evaluate the charge sums explicitly for \(n=1\) and \(n=2\). For
the single-sheet normalization there are only two insertions, with charges
\(\sigma,\tau=\pm1\) at \(\eta_-\) and \(\eta_+\). Since
\(\alpha+\lambda_{\mathfrak{a}}\bar\alpha=0\), the zero-mode selection rule imposes no
further constraint. Therefore
\begin{equation}
{\cal Z}_1
=
\sum_{\sigma,\tau=\pm1}
c_-(\sigma)c_+(\tau)
K_0^{\sigma\tau}.
\end{equation}
Using
\begin{equation}
c_+(+)=e^{-i\theta}\sqrt q,
\qquad
c_+(-)=\sqrt p,
\qquad
c_-(+)=\sqrt p,
\qquad
c_-(-)=e^{i\theta}\sqrt q,
\end{equation}
one obtains
\begin{equation}
\begin{aligned}
{\cal Z}_1
&=
pK_0^{-1}
+
qK_0^{-1}
+
\sqrt{pq}\,e^{-\ii\theta}K_0
+
\sqrt{pq}\,e^{\ii\theta}K_0  \\
&=
K_0^{-1}
+
2\sqrt{pq}\cos\theta\,K_0.
\end{aligned}
\end{equation}
Recall that for \(n=2\), the four insertions are
\begin{equation}
\xi_1=a,\qquad \xi_2=b,\qquad \xi_3=-a,\qquad \xi_4=-b,
\end{equation}
where \(\xi_1,\xi_3\) are incoming insertions and \(\xi_2,\xi_4\) are outgoing
insertions. The independent pairwise kernels are
\begin{equation}
K_{12}=K_{34}=K_-,
\qquad
K_{14}=K_{23}=K_+,
\qquad
K_{13}=K_{24}=H ,
\end{equation}
where \(K_{ij}\equiv K(\xi_i,\xi_j)\) and their explicit expressions are given in Eq.~(\ref{eq:basic-kernels}).

For a fixed charge sector
\((\sigma_1,\sigma_2,\sigma_3,\sigma_4)\), from Eq.~(\ref{eq:general-Zn}), it follows that the pairwise kernel factor is
\begin{equation}
K_-^{\sigma_1\sigma_2+\sigma_3\sigma_4}
K_+^{\sigma_1\sigma_4+\sigma_2\sigma_3}
H^{\sigma_1\sigma_3+\sigma_2\sigma_4}.
\end{equation}
Thus
\begin{equation}
{\cal Z}_2
=
\sum_{\sigma_1,\sigma_2,\sigma_3,\sigma_4=\pm1}
c_-(\sigma_1)c_+(\sigma_2)c_-(\sigma_3)c_+(\sigma_4)
K_-^{\sigma_1\sigma_2+\sigma_3\sigma_4}
K_+^{\sigma_1\sigma_4+\sigma_2\sigma_3}
H^{\sigma_1\sigma_3+\sigma_2\sigma_4}.
\end{equation}
The sixteen charge sectors fall into five groups. First, the two sectors
\((+,-,+,-)\) and \((-,+,-,+)\) give
\begin{equation}
(p^2+q^2)H^2K_-^{-2}K_+^{-2}
=
(p^2+q^2)H^2K_0^{-2}.
\end{equation}
Second, the two sectors \((+,+,+,+)\) and \((-,-,-,-)\) give
\begin{equation}
2pq\cos(2\theta)H^2K_-^2K_+^2
=
2pq\cos(2\theta)H^2K_0^2.
\end{equation}
Third, the two sectors \((+,-,-,+)\) and \((-,+,+,-)\) give
\begin{equation}
2pq\,H^{-2}K_-^{-2}K_+^2
=
2pq\,H^{-2}\frac{K_+^4}{K_0^2}.
\end{equation}
Fourth, the two sectors \((+,+,-,-)\) and \((-,-,+,+)\) give
\begin{equation}
2pq\,H^{-2}K_-^2K_+^{-2}
=
2pq\,H^{-2}\frac{K_-^4}{K_0^2}.
\end{equation}
Finally, the remaining eight sectors have zero net power of \(K_-\), \(K_+\),
and \(H\). Their coefficients sum to
\begin{equation}
4\sqrt{pq}\cos\theta .
\end{equation}
Combining these five contributions gives
\begin{equation}
{\cal Z}_2
=
H^2
\left[
(p^2+q^2)K_0^{-2}
+
2pq\cos(2\theta)K_0^2
\right]
+
2pq\,H^{-2}
\frac{K_-^4+K_+^4}{K_0^2}
+
4\sqrt{pq}\cos\theta .
\label{eq:Z2}
\end{equation}

Substituting Eqs.~\eqref{eq:v1qrefactor} and \eqref{eq:Z2} into \eqref{eq:general-Fn-vertex}, multiplying both the numerator and denominator by \(K_0^2\), and using
\({\cal J}_2{\cal R}_2=H^{-2}\), we obtain
\begin{equation}
F^{(2)}_{\Psi_2,+}(l,s,L;p,\theta)
=
\frac{
p^2+q^2
+2pq\cos(2\theta)K_0^4
+2pq\,\frac{K_-^4+K_+^4}{H^4}
+4\sqrt{pq}\cos\theta\,\frac{K_0^2}{H^2}
}{
\left(1+2\sqrt{pq}\cos\theta\,K_0^2\right)^2
}.
\label{eq:F2Psi2p}
\end{equation}
The second R\'enyi entropy shift is
\begin{equation}
S^{(\Psi_2,M)}_2(l,s,L;p,\theta)-S^{(0,M)}_2(l,s,L)
=
-\log F^{(2)}_{\Psi_2,+}(l,s,L;p,\theta).
\label{eq:entropy-shift}
\end{equation}
Two simple checks follow from \eqref{eq:F2Psi2p}. For \(p=1\) or \(p=0\), the state reduces to a single Umklapp excitation and one finds \(F^{(2)}_{V_{\pm},+}=1\). 
For an equal-weight superposition, \(p=q=1/2\), this reduces to
\begin{equation}
F^{(2)}_{\Psi_2,+}\bigg|_{p=1/2}
=
\frac{
\frac12\left(1+\cos(2\theta)K_0^4\right)
+\frac12\,\frac{K_-^4+K_+^4}{H^4}
+2\cos\theta\,\frac{K_0^2}{H^2}
}{
\left(1+\cos\theta\,K_0^2\right)^2
}.
\label{eq:equal-weight}
\end{equation}
\subsection{Unmeasured limit and phase sensitivity}
It is interesting to consider the unmeasured limit. As discussed in Sec.~\ref{subsec:unmeasured-limit}, this limit should not be obtained by
setting the slit length to zero directly. Instead, one should set \(s=\epsilon\) and then take \(\epsilon/L\to0\) with
\( x\equiv \frac{l}{L}\) fixed. In this limit the disk coordinates satisfy
\begin{equation}
\eta_-\to0,\qquad \eta_+\to0,\qquad
\frac{\eta_+}{\eta_-}\to e^{2\pi \ii x}.
\end{equation}
Equivalently, since \(a=\eta_-^{1/2}, b=\eta_+^{1/2}\), we have
\(a\to0,b\to0,\frac{b}{a}\to e^{\ii\pi x}\).

Using the definitions of \(K_\pm\), \(H\), and \(K_0\), one obtains
\begin{equation}
\frac{K_-}{H}\to
\frac{|1-e^{\ii\pi x}|^2}{4}
=
\sin^2\frac{\pi x}{2},
\end{equation}
and
\begin{equation}
\frac{K_+}{H}\to
\frac{|1+e^{\ii\pi x}|^2}{4}
=
\cos^2\frac{\pi x}{2}.
\end{equation}
On the other hand,
\begin{equation}
K_0\to0,\qquad
\frac{K_0^2}{H^2}\to0,\qquad
K_0^4\to0 .
\end{equation}
Therefore all terms in Eq.~\eqref{eq:F2Psi2p} that depend explicitly on the relative
phase \(\theta\) disappear in the ordinary-cylinder limit. We find
\begin{equation}
F^{(2)}_{\Psi_2}(x;p)
=
1-2pq\sin^2(\pi x)
+\frac{pq}{4}\sin^4(\pi x).
\end{equation}

For the equal-weight superposition \(p=q=1/2\), this reduces to
\begin{equation}
F^{(2)}_{\Psi_2}(x)
=
1-\frac{1}{2}\sin^2(\pi x)
+\frac{1}{16}\sin^4(\pi x)
=
\frac{1}{64}\left(7+\cos 2\pi x\right)^2 .
\end{equation}

This ordinary-cylinder result is independent of the relative phase
\(\theta\). By contrast, the finite-slit post-measurement expression in
Eq.~\eqref{eq:F2Psi2p} contains both \(\cos\theta\) and \(\cos 2\theta\) terms.
Thus the measurement-induced boundary activates phase-sensitive
interference sectors that are absent from the ordinary-cylinder
R\'enyi ratio. In this precise sense, the fixed-outcome measurement makes
the coherent relative phase visible to the post-measurement entanglement,
even though it is invisible in the unmeasured second R\'enyi entropy.
\section{Free-fermion numerics for the current excitation}
\label{sec:numerics-current}

We now describe the lattice method used to test the CFT prediction for the
post-measurement entanglement of the current excitation. The lattice model is
the critical XX chain at half filling. Since both the ground state and the
current excitation are Slater determinants, and since a fixed local
occupation-number measurement maps Slater determinants to Slater determinants,
the calculation can be performed entirely in terms of free-fermion correlation
matrices. 

\subsection{XX chain and its free-fermion representation}

We consider the periodic XX chain
\begin{equation}
    H_{\rm XX}
    =
    -\frac12
    \sum_{j=1}^{L}
    \left(
        \sigma^x_j\sigma^x_{j+1}
        +
        \sigma^y_j\sigma^y_{j+1}
    \right).
    \label{eq:xx-hamiltonian}
\end{equation}
Under the Jordan--Wigner transformation,
\begin{equation}
    c_j
    =
    \left(
        \prod_{m<j}\sigma^z_m
    \right)
    \frac{\sigma^x_j+\ii\sigma^y_j}{2},
    \label{eq:jw-transform}
\end{equation}
the Hamiltonian becomes a free-fermion hopping Hamiltonian. In the even
half-filled sector we use anti-periodic fermionic boundary conditions. The
single-particle momenta are then
\begin{equation}
    k_q=\frac{2\pi q}{L},
    \qquad
    q\in
    \left\{
        \pm\frac12,
        \pm\frac32,
        \dots,
        \pm\frac{L-1}{2}
    \right\}.
    \label{eq:apbc-momenta}
\end{equation}
The momentum-space creation operators are
\begin{equation}
    d_q^\dagger
    =
    \frac{1}{\sqrt L}
    \sum_{j=1}^{L}
    e^{\ii k_q j}c_j^\dagger .
\end{equation}

At half filling, the ground state is the Slater determinant
\begin{equation}
    |\Omega\rangle
    =
    \prod_{|q|<L/4}d_q^\dagger|0\rangle .
    \label{eq:xx-ground-state}
\end{equation}
Its one-body correlation matrix is
\begin{equation}
    C^{(0)}_{ij}
    =
    \langle\Omega|c_i^\dagger c_j|\Omega\rangle
    =
    \frac{1}{L}
    \sum_{|q|<L/4}
    e^{\ii k_q(j-i)}.
    \label{eq:ground-correlation}
\end{equation}

The CFT current \(J=\ii\partial\phi\) is represented on the lattice by the lowest right-moving particle-hole
excitation. Let \(q_R=\frac{L}{4}-\frac12\) be the highest occupied momentum just below the right Fermi point. We have
\begin{equation}
    |J\rangle
    =
    d_{q_R+1}^\dagger d_{q_R}|\Omega\rangle .
    \label{eq:lattice-current-state}
\end{equation}
This is again a Slater determinant. Its correlation matrix is obtained by
replacing the occupied momentum set \(\mathcal{K}_0=\{q:\ |q|<L/4\}\) of the ground state by
\begin{equation}
    \mathcal{K}_J
    =
    \left(
        \mathcal{K}_0\setminus\{q_R\}
    \right)
    \cup
    \{q_R+1\}.
    \label{eq:current-momentum-set}
\end{equation}
Thus
\begin{equation}
    C^{(J)}_{ij}
    =
    \frac{1}{L}
    \sum_{q\in \mathcal{K}_J}
    e^{\ii k_q(j-i)} .
    \label{eq:current-correlation}
\end{equation}

\subsection{Entanglement entropy from a correlation matrix}

Let \(|\Psi\rangle\) be a number-conserving Gaussian state with correlation
matrix
\begin{equation}
    C_{ij}
    =
    \langle\Psi|c_i^\dagger c_j|\Psi\rangle.
\end{equation}
For a subsystem \(B\), denote by \(C_B\) the restriction of \(C\) to sites in
\(B\):
\begin{equation}
    (C_B)_{ij}=C_{ij},
    \qquad
    i,j\in B.
\end{equation}
The reduced density matrix \(\rho_B\) is Gaussian and can be written as
\begin{equation}
    \rho_B
    =
    \frac{1}{Z_B}
    \exp\left(
        -\sum_{i,j\in B}h_{ij}c_i^\dagger c_j
    \right),
\end{equation}
where
\begin{equation}
    h=\log(C_B^{-1}-I),
\end{equation}
where \(I\) is the identity matrix. If \(\{\nu_\mu\}\) are the eigenvalues of \(C_B\), the R\'enyi entropy is
\begin{equation}
    S_n(B)
    =
    \frac{1}{1-n}
    \sum_\mu
    \log\left[
        \nu_\mu^n+(1-\nu_\mu)^n
    \right].
    \label{eq:renyi-from-corr}
\end{equation}

\subsection{Fixed measurement outcome as fermion-number conditioning}

We measure the spin variable \(\sigma^z\) on a spatial interval
\begin{equation}
    A=\{1,2,\dots,s\}.
\end{equation}
Under the Jordan--Wigner map, this is equivalent to measuring the local
fermion occupation numbers
\begin{equation}
    n_j=c_j^\dagger c_j,
    \qquad
    j\in A,
\end{equation}
up to a convention-dependent relabeling of occupied and empty sites.

A fixed measurement outcome is a binary string
\begin{equation}
    \bm m=(m_1,\dots,m_s),
    \qquad
    m_j\in\{0,1\}.
\end{equation}
Let
\begin{equation}
    O=\{j\in A:m_j=1\},
    \qquad
    E=\{j\in A:m_j=0\}
\end{equation}
be the occupied and empty measured sites. The corresponding projector is
\begin{equation}
    P_{\bm m}
    =
    \prod_{j\in O}n_j
    \prod_{j\in E}(1-n_j).
    \label{eq:measurement-projector}
\end{equation}
The unmeasured region is \(\bar A\).
For a Gaussian state \(|\Psi\rangle\), the normalized post-measurement state
on \(\bar A\) is
\begin{equation}
    \rho^{(\bm m)}_{\bar A}
    =
    \frac{
        \Tr_{A}
        \left(
            P_{\bm m}|\Psi\rangle\langle\Psi|P_{\bm m}
        \right)
    }{
        p_{\bm m}
    },
    \qquad
    p_{\bm m}
    =
    \langle\Psi|P_{\bm m}|\Psi\rangle .
    \label{eq:post-measurement-density}
\end{equation}
Because local occupation-number projection preserves fermionic Gaussianity,
\(\rho_{\bar A}^{(\bm m)}\) is again a number-conserving Gaussian state. It is
therefore fully determined by the conditional correlation matrix
\begin{equation}
    C^{(\bm m)}_{ij}
    =
    \frac{
        \langle\Psi|
        P_{\bm m}c_i^\dagger c_jP_{\bm m}
        |\Psi\rangle
    }{
        p_{\bm m}
    },
    \qquad
    i,j\in \bar A.
    \label{eq:conditional-corr-def}
\end{equation}

\subsection{Conditional correlation matrix}

The occupation statistics of a Slater determinant form a determinantal point
process with kernel \(C\). This gives an efficient formula for the conditional
kernel after fixing occupied and empty measured sites.

We first condition on the sites in \(O\) being occupied. Let \(\bar O=\{1,2,\dots,L\}\setminus O\).
The conditional kernel on \(\bar O\) is the Schur complement
\begin{equation}
    K^{(O)}
    =
    C_{\bar O\bar O}
    -
    C_{\bar OO}
    C_{OO}^{-1}
    C_{O\bar O}.
    \label{eq:condition-occupied}
\end{equation}
This formula follows directly from Wick's theorem. For \(i,j\notin O\),
\begin{equation}
    \frac{
        \left\langle
            c_i^\dagger c_j
            \prod_{a\in O} n_a
        \right\rangle
    }{
        \left\langle
            \prod_{a\in O} n_a
        \right\rangle
    }
    =
    C_{ij}
    -
    C_{iO}C_{OO}^{-1}C_{Oj}.
    \label{eq:occupied-schur}
\end{equation}
At this stage \(K^{(O)}\) is a matrix whose row and column indices both
belong to \(\bar O\). Since \(A=O\cup E\), we have \(\bar O = E\cup \bar A \).
We therefore regard \(K^{(O)}\) as a block matrix with respect to the
decomposition \(\bar O=E\cup\bar A\):
\begin{equation}
    K^{(O)}
    =
    \begin{pmatrix}
        K^{(O)}_{EE} & K^{(O)}_{E\bar A} \\
        K^{(O)}_{\bar A E} & K^{(O)}_{\bar A\bar A}
    \end{pmatrix}.
\end{equation}
Here, for example, \(K^{(O)}_{\bar A E}\) denotes the submatrix of
\(K^{(O)}\) with rows in \(\bar A\) and columns in \(E\).

We then condition on the sites in \(E\) being empty. Empty-site
conditioning for particles is equivalent to occupied-site conditioning
for holes, whose kernel is \(I-K^{(O)}\). Therefore the conditional
particle kernel on the unmeasured region \(\bar A\) is
\begin{equation}
    C^{(\bm m)}_{\bar A}
    =
    K^{(O)}_{\bar A\bar A}
    +
    K^{(O)}_{\bar A E}
    \left(I_E-K^{(O)}_{EE}\right)^{-1}
    K^{(O)}_{E\bar A}.
\end{equation}

\subsection{Subsystem entanglement after measurement}

For the half-filled XX chain, the natural fixed outcome in the \(\sigma^z\)
basis is the antiferromagnetic string  \(\bm m=(1,0,1,0,\dots)\),
or its translated partner \(\bm m=(0,1,0,1,\dots)\).

In fermion language this is an alternating occupation pattern on the measured
interval. In the bosonized low-energy theory this outcome corresponds to a
Neumann boundary condition for the compact boson field \(\phi=\varphi+\bar\varphi\), 
equivalently Dirichlet for its dual \(\varphi-\bar\varphi\). It is therefore
the natural lattice realization of the conformal boundary condition
\(\mathfrak a\) used in the fixed-outcome CFT calculation\footnote{In the convention commonly used in the XX-chain measurement literature,
the corresponding boundary condition is referred to as Dirichlet;
the Dirichlet field there is, up to normalization, our dual field
$\widetilde\phi=\varphi-\bar\varphi$. Thus it is equivalently Neumann
for $\phi=\varphi+\bar\varphi$ in the present convention.}
\cite{Rajabpour2015,Rajabpour2016,NajafiRajabpour2016}.

After measuring \(A=\{1,2,\dots,s\}\), the unmeasured region is \(\bar A=\{s+1,s+2,\dots,L\}\).
We choose the subsystem \(B\subset \bar A\) to be the first \(l\) sites adjacent
to the measured interval: \(B=\{s+1,s+2,\dots,s+l\}\).
This realizes the lattice version of the CFT geometry in which the interval
\(B\) starts at one endpoint of the slit.

\begin{figure}[t]
    \centering
    \subfloat[\(s/L=0.1\).]{
        \includegraphics[width=0.45\textwidth]{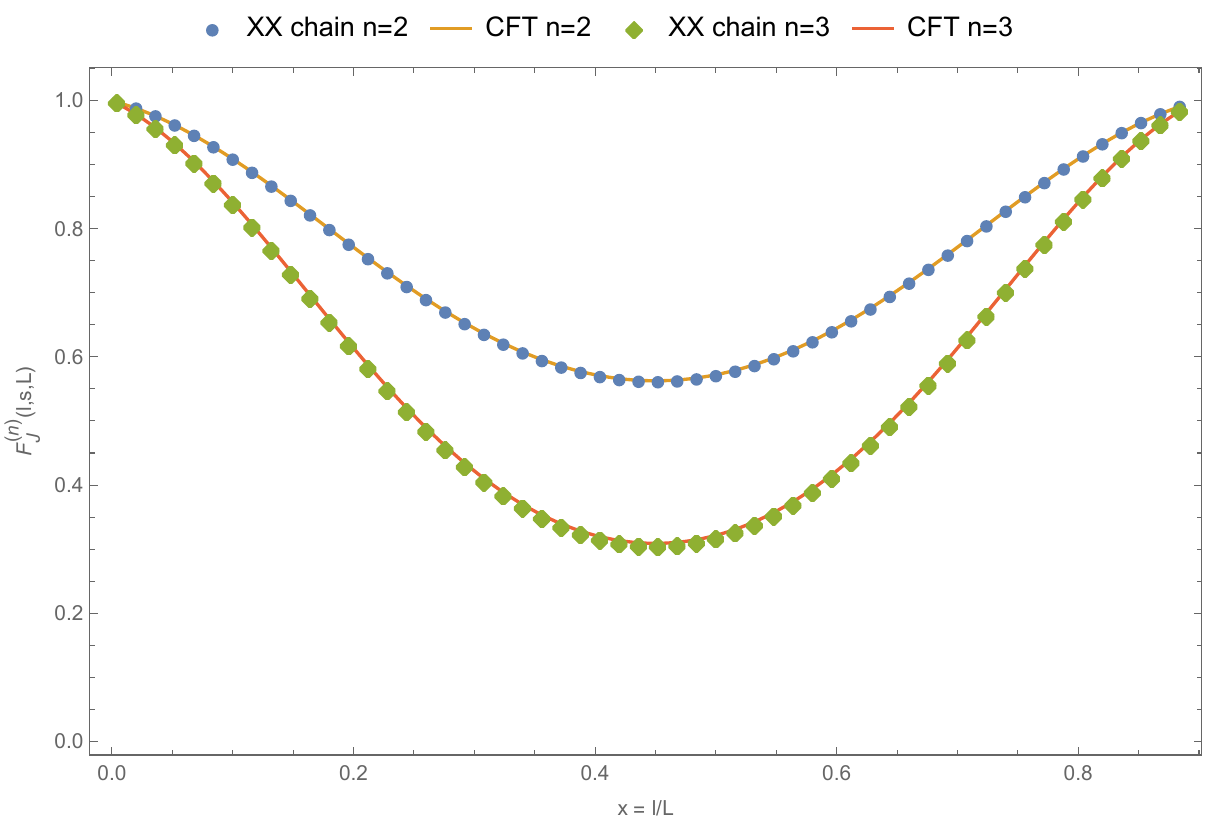}
        \label{fig:current-check-s01}
    }
    \hfill
    \subfloat[\(s/L=0.4\).]{
        \includegraphics[width=0.45\textwidth]{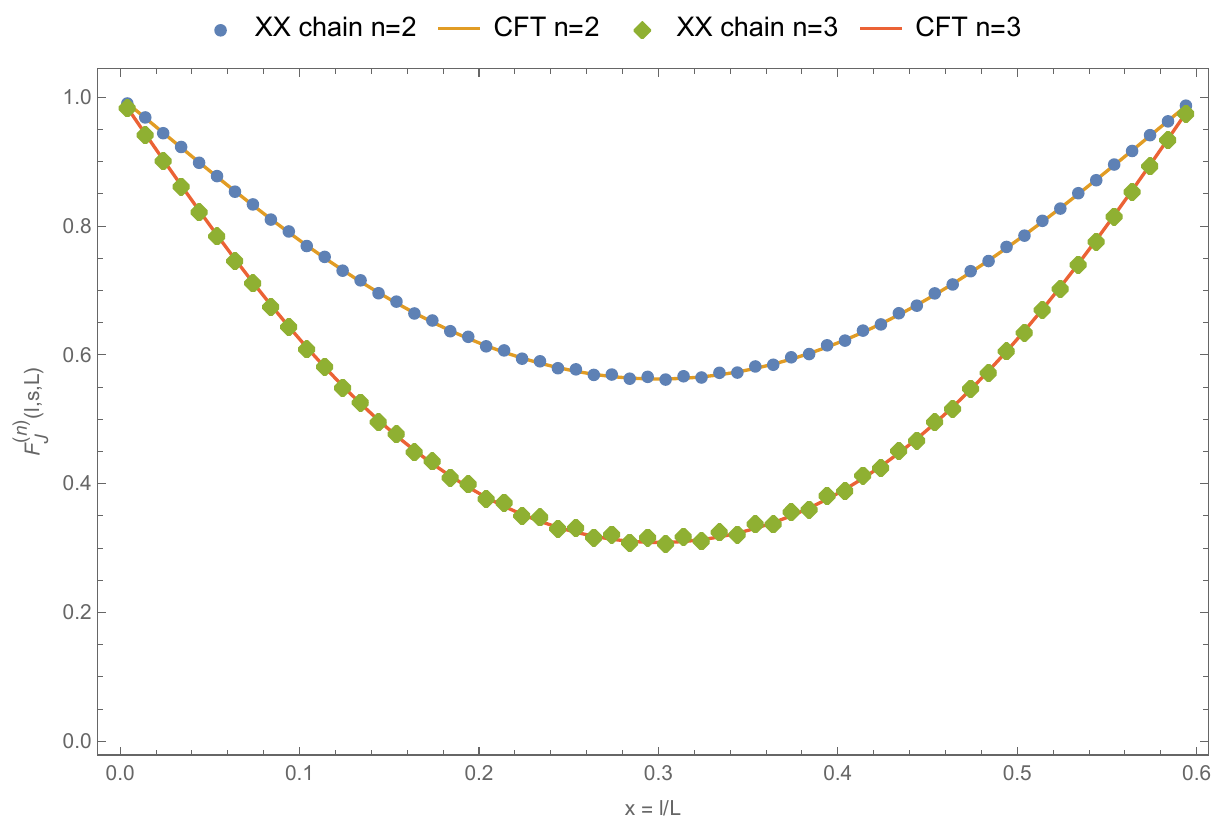}
        \label{fig:current-check-s02}
    }

    \caption{
    Numerical check of the post-measurement current-excitation ratios in the XX chain.
    The system size is \(L=500\).  The measured interval has length
    (a) \(s=50\), corresponding to \(s/L=0.1\), and
    (b) \(s=200\), corresponding to \(s/L=0.4\).
    The horizontal axis is \(x=l/L\).  Symbols denote lattice results obtained from the
    conditional free-fermion correlation matrix after projecting the measured sites onto
    a fixed alternating occupation pattern.  Solid curves are the CFT predictions for
    \(F_J^{(2)}\) (cf. Eq.~(\ref{eq:F2-current-rho})) and \(F_J^{(3)}\) (cf. Eq.~(\ref{eq:F3-current-rho})).  
    The agreements in both panels support the current
    hafnian formulas on the post-measurement slit geometry.
    }
    \label{fig:current-check}
\end{figure}

Given the conditional kernel \(C_{\bar A}^{(\bm m)}\), we restrict it to \(B\):
\begin{equation}
    C^{(\bm m)}_{B}
    =
    \left.
    C^{(\bm m)}_{\bar A}
    \right|_{B}.
\end{equation}
Let its eigenvalues be  \(\nu_1,\dots,\nu_l\). Then
\begin{equation}
    S^{(\Psi,M)}_n(l,s,L;\bm m)
    =
    \frac{1}{1-n}
    \sum_{\mu=1}^{l}
    \log\left[
        \nu_\mu^n+(1-\nu_\mu)^n
    \right].
    \label{eq:lattice-post-renyi}
\end{equation}
We test the current prediction for two different measurement lengths,
\(s/L=0.1\) and \(s/L=0.4\), while varying the subsystem size \(x=l/L\).
For each value of \(s\), we compare the lattice ratios
\begin{equation}
F_{J,\mathrm{lat}}^{(n)}
=
\frac{\mathrm{Tr}\rho_{J,B}^n}{\mathrm{Tr}\rho_{0,B}^n}
\end{equation}
with the CFT expressions for \(n=2\) (cf. Eq.~(\ref{eq:F2-current-rho})) and \(n=3\) (cf. Eq.~(\ref{eq:F3-current-rho})).  The two panels of
Fig.~\ref{fig:current-check} show that the agreement persists when the size of the
measured interval is changed.
\section{Multi-Slater determinant method for the superposition states}
\label{sec:numerics-superposition}
The current excitation discussed in Sec.~\ref{sec:current-excitation} can be
tested using a standard free-fermion correlation-matrix method, because both
the ground state and the current-excited state are single Slater determinants.
The left-right current superposition state in Sec.~\ref{sec:current-superposition}
and conjugate-vertex superposition state in Sec.~\ref{sec:vertex-superposition} require
a different method. Each component of the superposition is still a
Slater determinant, but their coherent sum is not a Gaussian state. Therefore
one cannot compute its entanglement from a single one-body correlation matrix.
Instead, for integer R\'enyi entropies one may use a multi-Slater determinant
method.

In this section we describe the multi-Slater determinant method, which gives a direct numerical
test of the \(n=2\) CFT prediction in Sec.~\ref{sec:current-superposition}
and Sec.~\ref{sec:vertex-superposition}.

\subsection{Lattice representatives of the superposition states}

We now specify the lattice Slater determinants used for the two coherent
superposition states studied in Sec.~\ref{sec:current-superposition}
and Sec.~\ref{sec:vertex-superposition}. Throughout this section we
work in the half-filled XX chain with anti-periodic fermionic boundary
conditions. The occupied momenta in the ground state form the Fermi sea
\begin{equation}
\mathcal{K}_0=\{q:\ |q|<L/4\}.
\end{equation}
We denote the right and left Fermi momenta just inside the filled sea by
\begin{equation}
q_R=\frac{L}{4}-\frac{1}{2},
\qquad
q_L=-\frac{L}{4}+\frac{1}{2}.
\end{equation}

The first superposition is the conjugate-current state discussed in
Sec.~\ref{sec:current-superposition}. Its two lattice components are the right- and left-moving
particle-hole excitations at the two Fermi points,
\begin{equation}
|J\rangle
=
d^\dagger_{q_R+1}d_{q_R}|\Omega\rangle,
\qquad
|\bar J\rangle
=
d^\dagger_{q_L-1}d_{q_L}|\Omega\rangle .
\end{equation}
They represent the CFT descendants \(J=\ii\partial\phi\) and \(\bar J=\ii\bar\partial\bar\phi\), 
respectively.

The lattice state corresponding to the coherent current superposition is
\begin{equation}
|\Psi_1\rangle
=
\sqrt p\,|J\rangle
+
e^{\ii\theta}\sqrt q\,|\bar J\rangle,
\qquad
q=1-p .
\end{equation}

The second superposition is the conjugate-vertex state discussed in
Sec.~\ref{sec:vertex-superposition}. Its two lattice components are the two Umklapp particle-hole
excitations
\begin{equation}
|V_+\rangle
=
d^\dagger_{q_R+1}d_{q_L}|\Omega\rangle,
\end{equation}
and
\begin{equation}
|V_-\rangle
=
d^\dagger_{q_L-1}d_{q_R}|\Omega\rangle .
\end{equation}
The corresponding coherent vertex superposition is
\begin{equation}
|\Psi_2\rangle
=
\sqrt p\,|V_+\rangle
+
e^{\ii\theta}\sqrt q\,|V_-\rangle,
\qquad
q=1-p .
\end{equation}

All four component states, \(|J\rangle, |\bar J\rangle, |V_+\rangle, |V_-\rangle,\)
are single Slater determinants. However, the coherent states
\(|\Psi_1\rangle\) and \(|\Psi_2\rangle\) are
linear combinations of two Slater determinants and are therefore not, in
general, Gaussian states. This is why the correlation-matrix method used
for a single current excitation must be replaced by the multi-Slater
determinant method described below.
\subsection{Slater determinant representation}
\label{subsec:slater-representation}
In the following we use a common notation for the two cases described above.
We write the two-component state as
\begin{equation}
|\Psi\rangle
=
\mathfrak{c}_1|\Phi_1\rangle+\mathfrak{c}_2|\Phi_2\rangle ,
\label{eq:lattice-superposition}
\end{equation}
where \(\mathfrak{c}_1=\sqrt p\) and \(\mathfrak{c}_2=e^{i\theta}\sqrt q\). For the current
superposition, \((|\Phi_1\rangle,|\Phi_2\rangle)=(|J\rangle,|\bar J\rangle)\);
for the vertex superposition,
\((|\Phi_1\rangle,|\Phi_2\rangle)=(|V_+\rangle,|V_-\rangle)\). Each
\(|\Phi_a\rangle\) is a Slater determinant, with orbital matrix denoted by
\(U^{(a)}\). Let \(U^{(a)}\) be its \(L\times N_f\) orbital matrix, where \(N_f=L/2\) at half
filling. Explicitly,
\begin{equation}
    U^{(a)}_{j\mu}
    =
    \frac{1}{\sqrt L}
    e^{\ii k_{q^{(a)}_\mu}j},
    \qquad
    j=1,\dots,L,
    \qquad
    \mu=1,\dots,N_f,
    \label{eq:orbital-matrix}
\end{equation}
where \(\{q^{(a)}_\mu\}\) is the occupied momentum set in
\(|\Phi_a\rangle\). The many-body state is
\begin{equation}
    |\Phi_a\rangle
    =
    \prod_{\mu=1}^{N_f}
    \left(
        \sum_{j=0}^{L-1}
        U^{(a)}_{j\mu}c_j^\dagger
    \right)|0\rangle .
\end{equation}
For an occupied configuration \(X=\{x_1,\dots,x_{N_f}\}\),
the wavefunction amplitude is
\begin{equation}
    \Phi_a(X)
    =
    \det U^{(a)}_X,
    \label{eq:slater-amplitude}
\end{equation}
where \(U^{(a)}_X\) is the \(N_f\times N_f\) submatrix obtained by restricting
\(U^{(a)}\) to the rows indexed by \(X\).

\subsection{Projection onto a fixed measurement outcome}
\label{subsec:project-slater}
We now describe how a single Slater determinant transforms under this
projection. After fixing the sites in \(O\) to be occupied, the post-measurement state on
\(\bar A\) has particle number \(N_f-|O|\). For a configuration \(Y\subset \bar A, |Y|=N_f-|O|\),
the projected amplitude is
\begin{equation}
   \Phi^{\,(\bm m)}_a(Y)
    =
    \det U^{(a)}_{O\cup Y}.
    \label{eq:projected-amplitude}
\end{equation}
The sites in \(E\) are automatically excluded because \(Y\subset \bar A\).

The wavefunction in Eq.~\eqref{eq:projected-amplitude} is again a Slater
determinant. To see this, define \(U^{(a)}_O\) as the \(|O|\times N_f\) matrix obtained by restricting \(U^{(a)}\) to the occupied
measured rows. Choose an \(N_f\times (N_f-|O|)\) matrix \(N_a\) whose columns span the
null space of \(U^{(a)}_O\):
\begin{equation}
    U^{(a)}_ON_a=0.
    \label{eq:nullspace-condition}
\end{equation}
Also choose an \(N_f\times |O|\) matrix \(Q_a\) such that \(U^{(a)}_OQ_a\) is
invertible, and define
\begin{equation}
    S_a=(Q_a,N_a).
\end{equation}
Then
\begin{equation}
    U^{(a)}_{O\cup Y}S_a
    =
    \begin{pmatrix}
        U^{(a)}_OQ_a & 0 \\
        U^{(a)}_YQ_a & U^{(a)}_YN_a
    \end{pmatrix}.
\end{equation}
Taking determinants gives
\begin{equation}
    \det U^{(a)}_{O\cup Y}
    =
    \kappa_a
    \det\left(U^{(a)}_YN_a\right),
    \label{eq:projection-determinant-factor}
\end{equation}
where
\begin{equation}
    \kappa_a
    =
    \frac{\det(U^{(a)}_OQ_a)}{\det S_a}.
    \label{eq:chi-factor}
\end{equation}
Therefore, up to the scalar factor \(\kappa_a\), the projected state is represented
on the unmeasured region \(\bar A\) by the \(|\bar A|\times (N_f-|O|)\) orbital
matrix \(U^{(a)}_{\bar A}N_a\).

We absorb the scalar factor \(\kappa_a\) into the unnormalized projected Slater
determinant and write
\begin{equation}
    P_{\bm m}|\Phi_a\rangle
    =
    | W^{(\bm m)}_a\rangle .
\end{equation}
Here \( W^{(\bm m)}_a\) denotes the orbital matrix of the unnormalized
Slater determinant obtained from the component \(|\Phi_a\rangle\) after
postselecting the measurement outcome \(\bm m\). The same construction applied
to the ground state gives
\begin{equation}
    P_{\bm m}|\Omega\rangle
    =
    | W^{(\bm m)}_0\rangle .
\end{equation}

\subsection{Post-measurement superposed state}

Applying the same fixed measurement outcome to the two-component state in
Eq.~(\ref{eq:lattice-superposition}) gives
\begin{equation}
    |\Psi^{(\bm m)}\rangle
    =
    P_{\bm m}|\Psi\rangle
    =
    \mathfrak c_1| W^{(\bm m)}_1\rangle
    +
    \mathfrak c_2| W^{(\bm m)}_2\rangle ,
    \qquad
    \mathfrak c_1=\sqrt p,\quad \mathfrak c_2=e^{\ii\theta}\sqrt q .
\end{equation}
Thus the fixed measurement maps each component to a Slater determinant, while
the superposed state remains a coherent sum of two generally non-orthogonal
Slater determinants.

Its norm is
\begin{equation}
    Z_\Psi
    =
    \langle \Psi^{(\bm m)}|
    \Psi^{(\bm m)}\rangle
    =
    \sum_{a,b=1}^{2}
    \mathfrak c_a \mathfrak c_b^\ast
    \det\!\left[
        \left( W^{(\bm m)}_b\right)^\dagger
         W^{(\bm m)}_a
    \right].
\end{equation}
For the projected ground state,
\begin{equation}
    Z_0
    =
    \langle  W^{(\bm m)}_0|
     W^{(\bm m)}_0\rangle
    =
    \det\!\left[
        \left( W^{(\bm m)}_0\right)^\dagger
         W^{(\bm m)}_0
    \right].
\end{equation}
The reduced density matrices and their purities are computed from these
unnormalized projected Slater determinants in the next subsection.
\begin{figure}[t]
    \centering
    \subfloat[\(F_{\Psi_1,+}^{(2)}\) as a function of \(x=l/L\), with
    \(L=500\), \(s=50\), \(p=1/2\), and \(\theta=\pi/4\).]{
        \includegraphics[width=0.48\textwidth]{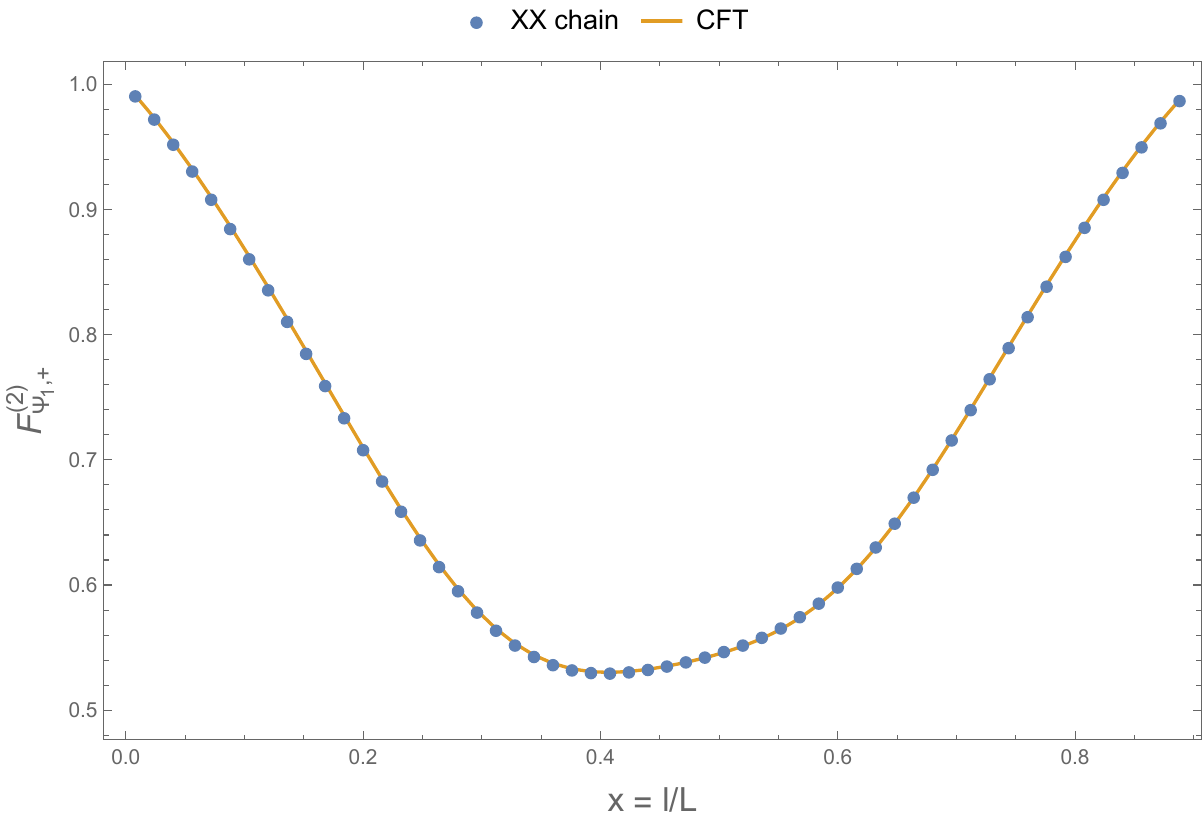}
        \label{fig:JJbar-length}
    }
    \hfill
    \subfloat[\(F_{\Psi_1,+}^{(2)}\) as a function of \(\theta\), with
    \(L=500\), \(s=50\), \(l=100\), and \(p=1/2\).]{
        \includegraphics[width=0.48\textwidth]{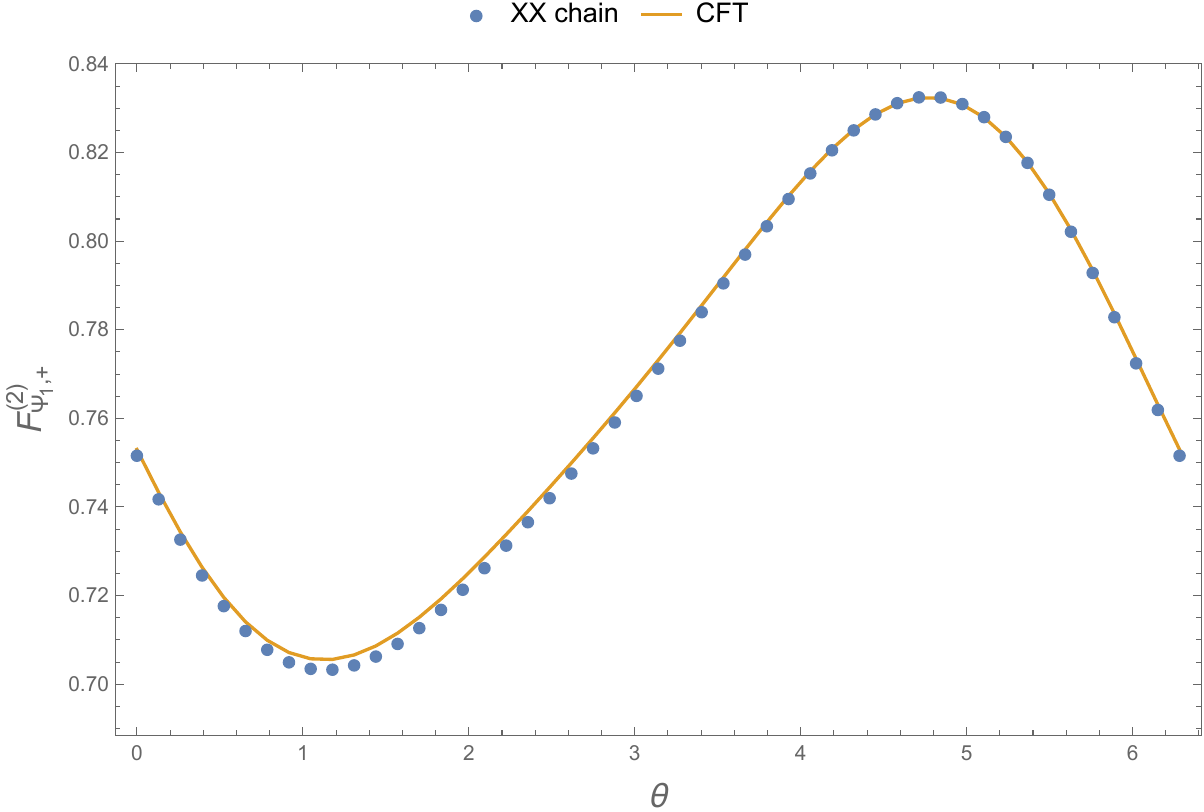}
        \label{fig:JJbar-phase}
    }

    \caption{
    Numerical check of the post-measurement second R\'enyi ratio for the
    \(J/\bar J\) superposition state.  Symbols denote XX-chain results obtained from
    the multi-Slater determinant calculation after projecting the measured sites onto a
    fixed alternating occupation pattern.  Solid curves are the CFT predictions from
    the disk four-current formula.  The left panel tests the dependence on the subsystem
    size \(x=l/L\), while the right panel tests the relative phase dependence.  The
    agreement in both panels verifies the coherent left-right interference terms induced
    by the measurement boundary.
    }
    \label{fig:JJbar-check}
\end{figure}
\subsection{Transition-Gaussian determinant formula for the purity}
\label{subsec:transition-gaussian-purity}
We now summarize the determinant formula used in the numerical evaluation.
The derivation is given in Appendix~\ref{app:transition-gaussian}. 

For two post-selected Slater determinants, define
\begin{equation}
S_{ba}^{(\bm m)}
=
\left(W_b^{(\bm m)}\right)^\dagger W_a^{(\bm m)},
\qquad
O_{ba}^{(\bm m)}
=
\det S_{ba}^{(\bm m)} .
\end{equation}
The transition correlation matrix restricted to the subsystem \(B\) is
\begin{equation}
G_B^{ab}
=
W_{a,B}^{(\bm m)}
\left(S_{ba}^{(\bm m)}\right)^{-1}
\left(W_{b,B}^{(\bm m)}\right)^\dagger .
\end{equation}
Here \(W_{a,B}^{(\bm m)}\) denotes the restriction of \(W_a^{(\bm m)}\) to the sites in
\(B\).

For the two-component post-selected state
\begin{equation}
|\Psi^{(\bm m)}\rangle
=
\mathfrak{c}_1 |W_1^{(\bm m)}\rangle
+
\mathfrak{c}_2 |W_2^{(\bm m)}\rangle ,
\end{equation}
the norm is
\begin{equation}
Z_\Psi
=
\sum_{a,b=1}^{2}
\mathfrak{c}_a \mathfrak{c}_b^*
O_{ba}^{(\bm m)} .
\end{equation}
The second R\'enyi purity of the normalized reduced density matrix is then
\begin{equation}
\operatorname{Tr}_B
\left[
\left(\rho_B^{\Psi,M}\right)^2
\right]
=
\frac{1}{Z_\Psi^2}
\sum_{a,b,c,d=1}^{2}
\mathfrak{c}_a \mathfrak{c}_b^* \mathfrak{c}_c \mathfrak{c}_d^*
O_{ba}^{(\bm m)}O_{dc}^{(\bm m)}
\det
\left[
\left(I_B-G_B^{ab}\right)
\left(I_B-G_B^{cd}\right)
+
G_B^{ab}G_B^{cd}
\right].
\end{equation}
This is the transition-Gaussian determinant formula used below.

For the post-selected ground state, the same formula reduces to the standard
free-fermion purity formula. Defining
\begin{equation}
G_B^{00}
=
W_{0,B}^{(\bm m)}
\left[
\left(W_0^{(\bm m)}\right)^\dagger W_0^{(\bm m)}
\right]^{-1}
\left(W_{0,B}^{(\bm m)}\right)^\dagger ,
\end{equation}
one obtains
\begin{equation}
\operatorname{Tr}_B
\left[
\left(\rho_B^{0,M}\right)^2
\right]
=
\det
\left[
\left(I_B-G_B^{00}\right)^2
+
\left(G_B^{00}\right)^2
\right].
\end{equation}
The universal quantity compared with the CFT prediction is therefore
\begin{equation}
F_{\rm num}^{(2)}(l,s,L)
=
\frac{
\operatorname{Tr}_B
\left[
\left(\rho_B^{\Psi,M}\right)^2
\right]
}{
\operatorname{Tr}_B
\left[
\left(\rho_B^{0,M}\right)^2
\right]
}.
\end{equation}
Equivalently,
\begin{equation}
F_{\rm num}^{(2)}
=
\exp
\left[
-
\left(
S_2^{\Psi,M}
-
S_2^{0,M}
\right)
\right].
\end{equation}
\subsection{Numerical test of the left-right current state}
We next test the phase-sensitive formula for the \(J/\bar J\) superposition state.
Unlike the single-current excitation, the post-measurement state is a coherent
superposition of two Slater determinants on the lattice.  We therefore evaluate the
second R\'enyi ratio using the transition-Gaussian determinant formula for multi-Slater
states. 

In Fig.~\ref{fig:JJbar-check}, we show two complementary checks.  In the left panel we
fix \(p=1/2\) and \(\theta=\pi/4\), and vary the subsystem size \(x=l/L\).  This tests
the geometric dependence of the disk four-point function.  In the right panel we fix
\(l=100\) and vary the relative phase \(\theta\), which directly probes the coherent
interference between the \(J\) and \(\bar J\) sectors.  The agreement in both panels
confirms the \(J/\bar J\) superposition formula \eqref{eq:Jbar-n2-F2}, including its phase-dependent
interference terms.
\subsection{Numerical test of the conjugate-vertex state}
We finally test the CFT formula for the conjugate-vertex superposition.  This
example provides a check of the vertex-operator interference terms in the
replica correlator.  We consider the lattice state
\begin{equation}
|\Psi_{2}\rangle
=
\sqrt{p}\,|V_+\rangle
+
e^{\ii\theta_{\rm lat}}\sqrt{1-p}\,|V_-\rangle .
\end{equation}
The two components \(V_+\) and \(V_-\) have identical conformal weights, and
therefore the first conformal map does not generate a relative
holomorphic-antiholomorphic Jacobian phase, unlike in the \(J/\bar J\) case.
However, the phase appearing in the CFT vertex correlator is defined with
respect to the cocycle and square-root branch convention used in the disk
uniformization.  The lattice Slater representatives carry their own phase
convention.

We fix this freedom after the measurement projection by requiring
\begin{equation}
\langle W_-|W_+\rangle \in \mathbb{R}_{>0}.
\end{equation}
With this convention, the phase entering the CFT formula is related to the
lattice phase by
\begin{equation}
\theta_{\rm CFT}
=
\theta_{\rm lat}
-\frac{\pi l}{L}
+\frac{\pi s}{4L}.
\label{eq:vertex-phase-matching}
\end{equation}
Here $\theta_{\rm CFT}$ should be understood as the effective relative phase $\theta_{\rm eff}$ defined in Sec.~\ref{subsec:v11-n2}, which includes the
boundary zero-mode shift $2\chi_a$ and is subsequently denoted simply
by $\theta$. Thus, a fixed value of $\chi_{\mathfrak{a}}$ is absorbed into the
CFT--lattice phase matching in Eq.~\eqref{eq:vertex-phase-matching}. This is a branch-convention matching phase in the replica geometry, not an
additional dynamical phase.

We should emphasize that Eq.~\eqref{eq:vertex-phase-matching} is used here as an empirically identified branch-matching prescription suggested by the lattice data. A first-principles derivation would require tracking the cocycle phases, square-root branch choices, and the phases of the transition determinants in each interference sector of the multi-Slater calculation. Since this phase bookkeeping is technically involved and does not affect the universal amplitude structure tested below, we leave such an analytic derivation for future work.

\begin{figure}[t]
\centering
    \subfloat[\(F_{\Psi_2,+}^{(2)}\) as a function of \(x=l/L\), with
    \(L=800\), \(s=100\), \(p=1/2\), and \(\theta=\pi/4\).]{
        \includegraphics[width=0.48\textwidth]{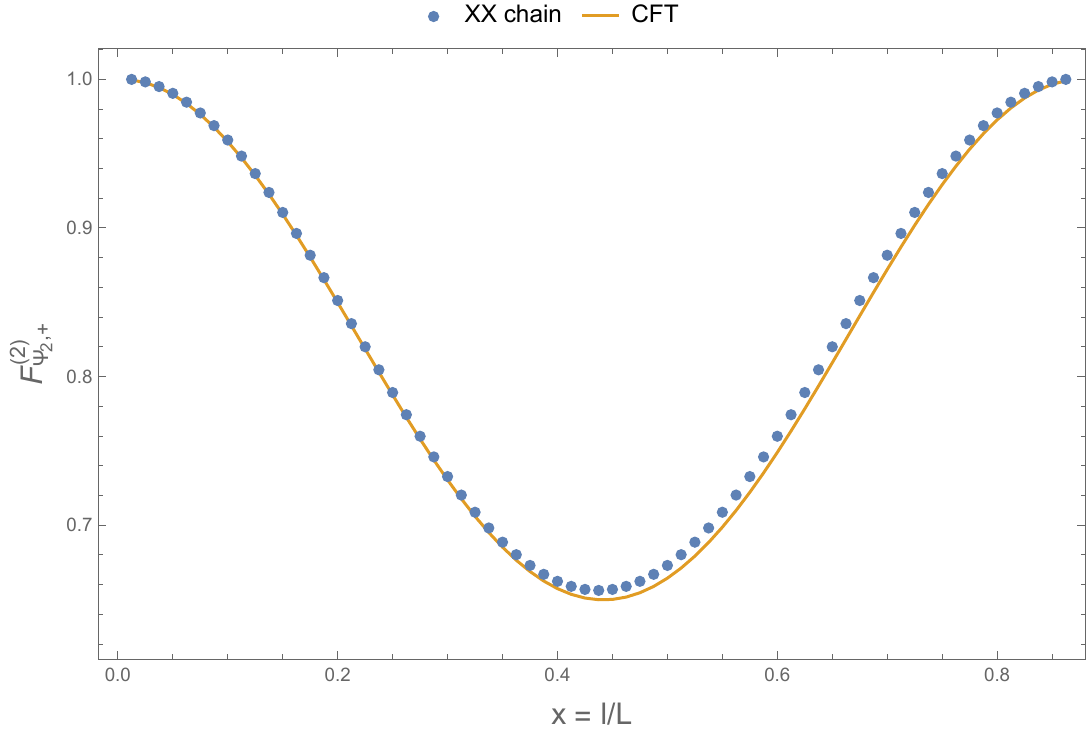}
        \label{fig:vertex-length}
    }
    \hfill
    \subfloat[\(F_{\Psi_2,+}^{(2)}\) as a function of \(\theta\), with
    \(L=800\), \(s=200\), \(l=100\), and \(p=1/2\).]{
        \includegraphics[width=0.48\textwidth]{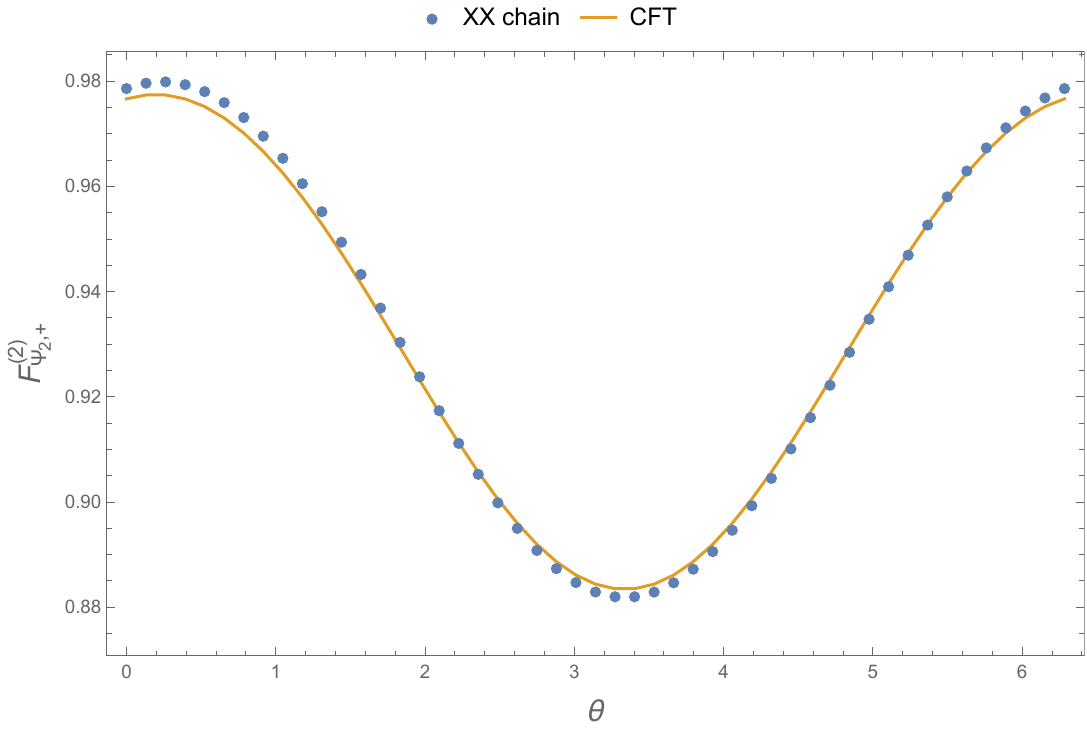}
        \label{fig:vertex-phase}
    }
\caption{
  Numerical check of the conjugate-vertex superposition. The blue points are
  the XX-chain results, and the orange curves are the CFT predictions.  In the
  phase scan, the CFT curve is evaluated using the branch-matched phase
  \(\theta_{\rm CFT}=\theta_{\rm lat}-\pi l/L+\pi s/(4L)\). The agreement shows
  that the CFT expression captures both the spatial dependence and the
  phase-sensitive interference structure of the vertex superposition. The
  small residual deviations are attributed to finite-size corrections.
  }
\label{fig:vertex-check}
\end{figure}

In Fig.~\ref{fig:vertex-check}, we show two complementary tests.  The left
panel varies \(x=l/L\) at fixed \(p\) and \(\theta_{\rm lat}\), testing the
spatial dependence of the CFT expression.  The right panel varies
\(\theta_{\rm lat}\) at fixed \(l\), testing the phase-sensitive interference
terms.  After the phase matching in Eq.~\eqref{eq:vertex-phase-matching}, the
CFT prediction in Eq.~\eqref{eq:F2Psi2p} captures both the spatial dependence and the phase dependence of
the lattice data.  The remaining deviations are small finite-size corrections.

\section{Conclusion}
\label{sec:conclusion}

In this paper, we compute the entanglement of low-energy excited states after a local measurement for post-selected outcomes whose induced boundary
perturbation flows in the infrared to a conformal boundary condition. The measurement interval is represented by a slit carrying a conformal boundary condition, while the excited state is implemented by operator insertions in the Euclidean path integral. After mapping the slit cylinder to the upper half-plane and uniformizing the replicated geometry, the excited-state correction is reduced to a normalized disk correlation function. This formulation separates the universal contribution of the measurement geometry from the operator-dependent contribution of the excitation.

We applied this construction to the compact free boson CFT. For the chiral current excitation, the relevant disk correlators are current hafnians, leading to closed expressions for the second and third R\'enyi ratios. We also analyzed coherent superpositions. In the \(J/\bar J\) case, the measurement-induced boundary makes the relative phase observable by mixing holomorphic and antiholomorphic sectors. For conjugate compact vertex operators, the single-component excitations are trivial in the normalized ratio considered here, but their coherent superposition produces a nontrivial contribution from interference sectors across replicas. We further showed that the corresponding ordinary-cylinder second R\'enyi ratio is independent of the relative phase, whereas the finite-slit post-measurement ratio contains explicit phase-sensitive interference terms. Thus the fixed-outcome measurement reveals coherent phase information that is invisible to the unmeasured second R\'enyi entropy.

We further described lattice methods for testing these predictions in the critical XX chain. The current excitation can be treated by conditional free-fermion correlation matrices, whereas coherent superpositions require a multi-Slater determinant method because the post-measurement state is no longer Gaussian. The transition-Gaussian determinant formula gives an efficient way to compute integer R\'enyi entropies for such superposed Slater states.

Several directions remain open. It would be quite interesting to extend the present fixed-outcome construction to Born-averaged measurement-induced entanglement, where additional replicas and compact-boson winding sectors are expected to play an essential role \cite{KhannaVasseur2026}. More broadly, one could study how measurements affect quench dynamics. The numerical framework developed here may also be useful for non-equilibrium problems, such as crosscap quench dynamics \cite{Zhang:2024rnh, WeiYoneta2026, Chalas:2024yts, Chen:2025wzd}, where superpositions of Gaussian states arise naturally.
\section*{Acknowledgments}
This work was supported  by the National Natural Science Foundation of China, Grant No.\ 12005081 and No.\ 12465014.
\appendix

\section{Transition Gaussian Density Matrices for Multi-Slater States}
\label{app:transition-gaussian}

In this appendix we derive the transition correlation matrix and the determinant contraction formula used in the multi-Slater determinant evaluation of the second R\'enyi entropy. The derivation relies on three standard ingredients: the overlap formula for non-orthogonal Slater determinants \cite{Lowdin1955}, the generalized Wick theorem for non-orthogonal fermionic product states \cite{BalianBrezin1969,Burton2021}, and the fermionic Gaussian trace identity, often referred to as the Klich formula \cite{Klich2003}. The resulting determinant expressions are closely related to the correlation-matrix approach to free-fermion entanglement \cite{Peschel2003,PeschelEisler2009} and to determinant formulas used in auxiliary-field and multi-Slater calculations of R\'enyi entropies \cite{Grover2013,AssaadLangToldin2014,Assaad2015}.

In the application to the post-measurement problem in Sec.~\ref{sec:numerics-superposition}, the matrices
\(W_a\) below are the post-selected matrices \(W_a^{(\bm m)}\). We suppress the
superscript \((\bm m)\) in this appendix. They are allowed to be unnormalized.

The unmeasured region after the projective measurement is \(\bar A\), and let \(B\subset \bar A\) be the subsystem whose reduced density matrix is required. We write the complement of \(B\) inside \(\bar A\) as \(C=\bar A\setminus B\). Consider a set of number-conserving Slater determinants
\begin{equation}
    |W_a\rangle
    =
    \prod_{\mu=1}^{N_f}
    \left(
        \sum_{i\in \bar A} (W_a)_{i\mu} c_i^\dagger
    \right)|0\rangle ,
\end{equation}
where \(W_a\) is an \(|\bar A|\times N_f\) matrix whose columns are the occupied single-particle orbitals of the Slater determinant labelled by \(a\). The orbitals in different determinants need not be mutually orthogonal. For two such determinants one has
\begin{equation}
    \langle W_b|W_a\rangle
    =
    \det S_{ba},
    \qquad
    S_{ba}=W_b^\dagger W_a .
\end{equation}
We denote this overlap by
\begin{equation}
    O_{ba}\equiv \det(W_b^\dagger W_a).
\end{equation}

We first derive the reduced transition operator
\begin{equation}
    \rho_B^{ab}
    \equiv
    \mathrm{Tr}_{C}\, |W_a\rangle\langle W_b| .
\end{equation}
Although \(\rho_B^{ab}\) is not necessarily Hermitian when \(a\neq b\), it is still a Gaussian transition operator. Its kernel is determined by a transition correlation matrix.

Introduce a one-body source \(X_B\) acting only on the subsystem \(B\), and extend it to the full region \(\bar A\) by
\begin{equation}
    X_{\bar A} =
    \begin{pmatrix}
        X_B & 0 \\
        0 & 0
    \end{pmatrix},
\end{equation}
where the block decomposition is with respect to \(\bar A=B\cup C\). The standard Slater-determinant identity gives
\begin{equation}
    \langle W_b|
    \exp(c_{\bar A}^\dagger X_{\bar A} c_{\bar A})
    |W_a\rangle
    =
    \det\!\left(W_b^\dagger e^{X_{\bar A}}W_a\right).
\end{equation}
Since the source acts only on \(B\), we may write
\begin{equation}
    e^{X_{\bar A}}
    =
    I_{\bar A}+P_B(e^{X_B}-I_B)P_B ,
\end{equation}
where \(P_B\) is the projector from \(\bar A\) to \(B\). Hence
\begin{equation}
\begin{aligned}
    W_b^\dagger e^{X_{\bar A}}W_a
    &=
    W_b^\dagger W_a
    +
    W_{b,B}^\dagger (e^{X_B}-I_B)W_{a,B}      \\
    &=
    S_{ba}
    +
    W_{b,B}^\dagger (e^{X_B}-I_B)W_{a,B}.
\end{aligned}
\end{equation}
Since \(X_{\bar A}\) is only non-vanishing on \(B\), we have \(c_{\bar A}^\dagger X_{\bar A} c_{\bar A}=c_B^\dagger X_B c_B\).
Taking the determinant and factoring out \(S_{ba}\), one obtains
\begin{equation}
\begin{aligned}
    \langle W_b|
    \exp(c_B^\dagger X_B c_B)
    |W_a\rangle
    &=
    \det S_{ba}\,
    \det\!\left[
        I_{N_f}+
        S_{ba}^{-1}
        W_{b,B}^\dagger
        (e^{X_B}-I_B)
        W_{a,B}
    \right] .
\end{aligned}
\end{equation}
Using Sylvester's determinant identity,
\begin{equation}
    \det(I+AB)=\det(I+BA),
\end{equation}
this becomes
\begin{equation}
\begin{aligned}
    \langle W_b|
    \exp(c_B^\dagger X_B c_B)
    |W_a\rangle
    &=
    O_{ba}\,
    \det\!\left[
        I_B+
        (e^{X_B}-I_B)
        W_{a,B}
        S_{ba}^{-1}
        W_{b,B}^\dagger
    \right]                                      \\
    &=
    O_{ba}\,
    \det\!\left[
        I_B-G_B^{ab}+e^{X_B}G_B^{ab}
    \right],
\end{aligned}
\end{equation}
where we have defined
\begin{equation}
    G_B^{ab}
    =
    W_{a,B}
    (W_b^\dagger W_a)^{-1}
    W_{b,B}^\dagger .
    \label{eq:transition-G}
\end{equation}
This is the transition correlation matrix used in the main text. For \(a=b\), the transition kernel becomes the ordinary one-body correlation
matrix of the normalized Slater state associated with \(W_a\). If, in addition,
the columns of \(W_a\) are orthonormal, then Eq.~\eqref{eq:transition-G} reduces to
\begin{equation}
G_B^{aa}=W_{a,B}W_{a,B}^\dagger .
\end{equation}
For unnormalized post-selected Slater determinants, the more general expression \eqref{eq:transition-G} should be used.
For \(a\neq b\), \(G_B^{ab}\) is generally non-Hermitian because it is associated with the transition operator
\(|W_a\rangle\langle W_b|\), rather than with a physical density matrix. Nevertheless, it plays the role of the Wick contraction kernel: normalized matrix elements between \(\langle W_b|\) and \(|W_a\rangle\) are evaluated by the generalized Wick theorem with contraction matrix \(G_B^{ab}\)\cite{BalianBrezin1969,Burton2021}.

The previous generating function shows that the reduced transition operator can be written as
\begin{equation}
    \rho_B^{ab}
    =
    O_{ba}\,\widehat{\rho}_B(G_B^{ab}),
    \label{eq:transition-rho}
\end{equation}
where \(\widehat{\rho}_B(G)\) is a normalized Gaussian operator satisfying
\begin{equation}
    \mathrm{Tr}_B\,\widehat{\rho}_B(G)=1,
\end{equation}
and
\begin{equation}
    \mathrm{Tr}_B
    \left[
        \widehat{\rho}_B(G)
        \exp(c_B^\dagger X_B c_B)
    \right]
    =
    \det\!\left[
        I_B-G+e^{X_B}G
    \right].
    \label{eq:gaussian-generating}
\end{equation}
Equivalently, when \(I_B-G\) is invertible, one may write
\begin{equation}
    \widehat{\rho}_B(G)
    =
    \det(I_B-G)
    \exp\!\left[
        c_B^\dagger
        \log\!\left(G(I_B-G)^{-1}\right)
        c_B
    \right].
    \label{eq:gaussian-density}
\end{equation}
Equation~\eqref{eq:gaussian-density} is the usual free-fermion reduced-density-matrix form, analytically continued to a possibly non-Hermitian transition kernel.

We now derive the determinant formula for the contraction of two transition reduced density matrices. Let \(G_1\) and \(G_2\) be two such kernels. Define
\begin{equation}
    e^{K_{G_1}}=G_1(I_B-G_1)^{-1},
    \qquad
    e^{K_{G_2}}=G_2(I_B-G_2)^{-1}.
\end{equation}
Using Eq.~\eqref{eq:gaussian-density}, we have
\begin{equation}
\begin{aligned}
    \mathrm{Tr}_B
    \left[
        \widehat{\rho}_B(G_1)\widehat{\rho}_B(G_2)
    \right]
    &=
    \det(I_B-G_1)\det(I_B-G_2)                         \\
    &\quad \times
    \mathrm{Tr}_B
    \left[
        e^{c_B^\dagger K_{G_1} c_B}
        e^{c_B^\dagger K_{G_2} c_B}
    \right].
\end{aligned}
\end{equation}
The fermionic Gaussian trace identity \cite{Klich2003} gives
\begin{equation}
    \mathrm{Tr}_B
    \left[
        e^{c_B^\dagger K_{G_1} c_B}
        e^{c_B^\dagger K_{G_2} c_B}
    \right]
    =
    \det\!\left(I_B+e^{K_{G_1}}e^{K_{G_2}}\right).
    \label{eq:klich}
\end{equation}
Substituting \(e^{K_{G_1}}=G_1(I_B-G_1)^{-1}\) and \(e^{K_{G_2}}=G_2(I_B-G_2)^{-1}\), we find
\begin{equation}
\begin{aligned}
    \mathrm{Tr}_B
    \left[
        \widehat{\rho}_B(G_1)\widehat{\rho}_B(G_2)
    \right]
    &=
    \det(I_B-G_1)\det(I_B-G_2)                         \\
    &\quad \times
    \det\!\left[
        I_B+
        G_1(I_B-G_1)^{-1}
        G_2(I_B-G_2)^{-1}
    \right].
\end{aligned}
\end{equation}
Using determinant cyclicity and multiplying the factors inside the determinant, this becomes
\begin{equation}
    \mathrm{Tr}_B
    \left[
        \widehat{\rho}_B(G_1)\widehat{\rho}_B(G_2)
    \right]
    =
    \det\!\left[
        (I_B-G_1)(I_B-G_2)+G_1G_2
    \right].
    \label{eq:gaussian-contraction}
\end{equation}

Finally, applying Eq.~\eqref{eq:gaussian-contraction} to the transition operators in Eq.~\eqref{eq:transition-rho}, with
\begin{equation}
    G_1=G_B^{ab},
    \qquad
    G_2=G_B^{cd},
\end{equation}
we obtain
\begin{equation}
\begin{aligned}
    &\mathrm{Tr}_B
    \left[
        \mathrm{Tr}_{C}|W_a\rangle\langle W_b|\,
        \mathrm{Tr}_{C}|W_c\rangle\langle W_d|
    \right]                                                \\
    &\qquad =
    O_{ba}O_{dc}\,
    \det\!\left[
        (I_B-G_B^{ab})(I_B-G_B^{cd})
        +
        G_B^{ab}G_B^{cd}
    \right].
\end{aligned}
\label{eq:transition-contraction}
\end{equation}
This is the desired contraction identity. 

For a single Slater determinant, the reduced transition operator
\(\rho_B^{aa}\) has trace \(O_{aa}\). The normalized reduced density matrix is
therefore
\begin{equation}
\widehat \rho_B^{(a)}
=
\frac{\rho_B^{aa}}{O_{aa}} .
\end{equation}
Using Eq.~\eqref{eq:transition-contraction} with \(a=b=c=d\), one obtains
\begin{equation}
\operatorname{Tr}_B
\left[
\left(\widehat \rho_B^{(a)}\right)^2
\right]
=
\det\left[
(I_B-G_B^{aa})^2+(G_B^{aa})^2
\right].
\end{equation}
If \(W_a^\dagger W_a=I\), then \(O_{aa}=1\) and
\(G_B^{aa}=W_{a,B}W_{a,B}^\dagger\), recovering the standard free-fermion
purity formula.
For a coherent superposition of Slater determinants,
\begin{equation}
    |\Psi\rangle
    =
    \sum_a \mathfrak{c}_a |W_a\rangle ,
\end{equation}
the reduced density matrix on \(B\) is
\begin{equation}
    \rho_B
    =
    \sum_{a,b}
    \mathfrak{c}_a \mathfrak{c}_b^*
    \mathrm{Tr}_{C}|W_a\rangle\langle W_b| .
\end{equation}
Therefore the second R\'enyi trace is obtained by summing Eq.~\eqref{eq:transition-contraction} over all transition sectors:
\begin{equation}
\begin{aligned}
    \mathrm{Tr}_B\,\rho_B^2
    &=
    \sum_{a,b,c,d}
    \mathfrak{c}_a \mathfrak{c}_b^*
    \mathfrak{c}_c \mathfrak{c}_d^*
    O_{ba}O_{dc}                                      \\
    &\quad \times
    \det\!\left[
        (I_B-G_B^{ab})(I_B-G_B^{cd})
        +
        G_B^{ab}G_B^{cd}
    \right].
\end{aligned}
\end{equation}
This expression is the form used in the multi-Slater determinant evaluation of the second R\'enyi entropy.


\begin{thebibliography}{99}

\bibitem{Holzhey1994}
C.~Holzhey, F.~Larsen, and F.~Wilczek,
``Geometric and renormalized entropy in conformal field theory,''
\href{https://doi.org/10.1016/0550-3213(94)90402-2}
{\emph{Nucl. Phys. B} \textbf{424}, 443--467 (1994)}.

\bibitem{CalabreseCardy2004}
P.~Calabrese and J.~Cardy,
``Entanglement entropy and quantum field theory,''
\href{https://doi.org/10.1088/1742-5468/2004/06/P06002}
{\emph{J. Stat. Mech.: Theory Exp.} \textbf{2004}, P06002 (2004)}.

\bibitem{CalabreseCardy2009}
P.~Calabrese and J.~Cardy,
``Entanglement entropy and conformal field theory,''
\href{https://doi.org/10.1088/1751-8113/42/50/504005}
{\emph{J. Phys. A: Math. Theor.} \textbf{42}, 504005 (2009)}.

\bibitem{DiFrancesco1997}
P.~Di~Francesco, P.~Mathieu, and D.~S\'en\'echal,
\href{https://doi.org/10.1007/978-1-4612-2256-9}
{\emph{Conformal Field Theory}. Springer, New York (1997)}.

\bibitem{Cardy1984}
J.~L. Cardy,
``Conformal invariance and surface critical behavior,''
\href{https://doi.org/10.1016/0550-3213(84)90241-4}
{\emph{Nucl. Phys. B} \textbf{240}, 514--532 (1984)}.

\bibitem{Cardy2004BCFT}
J.~Cardy,
``Boundary conformal field theory,''
\href{https://doi.org/10.1016/B0-12-512666-2/00398-9}
{in \emph{Encyclopedia of Mathematical Physics},
J.-P.~Fran\c{c}oise, G.~L.~Naber, and T.~S.~Tsun, eds.,
pp.~333--340. Elsevier (2006)}.

\bibitem{AlcarazBerganzaSierra2011}
F.~C. Alcaraz, M.~I. Berganza, and G.~Sierra,
``Entanglement of low-energy excitations in conformal field theory,''
\href{https://doi.org/10.1103/PhysRevLett.106.201601}
{\emph{Phys. Rev. Lett.} \textbf{106}, 201601 (2011)}.

\bibitem{BerganzaAlcarazSierra2012}
M.~I. Berganza, F.~C. Alcaraz, and G.~Sierra,
``Entanglement of excited states in critical spin chains,''
\href{https://doi.org/10.1088/1742-5468/2012/01/P01016}
{\emph{J. Stat. Mech.: Theory Exp.} \textbf{2012}, P01016 (2012)}.

\bibitem{LiChenFisher2018}
Y.~Li, X.~Chen, and M.~P.~A. Fisher,
``Quantum Zeno effect and the many-body entanglement transition,''
\href{https://doi.org/10.1103/PhysRevB.98.205136}
{\emph{Phys. Rev. B} \textbf{98}, 205136 (2018)}.

\bibitem{SkinnerRuhmanNahum2019}
B.~Skinner, J.~Ruhman, and A.~Nahum,
``Measurement-induced phase transitions in the dynamics of entanglement,''
\href{https://doi.org/10.1103/PhysRevX.9.031009}
{\emph{Phys. Rev. X} \textbf{9}, 031009 (2019)}.

\bibitem{JianYouVasseurLudwig2020}
C.-M. Jian, Y.-Z. You, R.~Vasseur, and A.~W.~W. Ludwig,
``Measurement-induced criticality in random quantum circuits,''
\href{https://doi.org/10.1103/PhysRevB.101.104302}
{\emph{Phys. Rev. B} \textbf{101}, 104302 (2020)}.

\bibitem{BaoChoiAltman2020}
Y.~Bao, S.~Choi, and E.~Altman,
``Theory of the phase transition in random unitary circuits with measurements,''
\href{https://doi.org/10.1103/PhysRevB.101.104301}
{\emph{Phys. Rev. B} \textbf{101}, 104301 (2020)}.

\bibitem{LiChenLudwigFisher2021}
Y.~Li, X.~Chen, A.~W.~W. Ludwig, and M.~P.~A. Fisher,
``Conformal invariance and quantum nonlocality in critical hybrid circuits,''
\href{https://doi.org/10.1103/PhysRevB.104.104305}
{\emph{Phys. Rev. B} \textbf{104}, 104305 (2021)}.

\bibitem{ZabaloGullansWilsonVasseurLudwigGopalakrishnanHusePixley2022}
A.~Zabalo, M.~J. Gullans, J.~H. Wilson, R.~Vasseur, A.~W.~W. Ludwig,
S.~Gopalakrishnan, D.~A. Huse, and J.~H. Pixley,
``Operator scaling dimensions and multifractality at measurement-induced transitions,''
\href{https://doi.org/10.1103/PhysRevLett.128.050602}
{\emph{Phys. Rev. Lett.} \textbf{128}, 050602 (2022)}.

\bibitem{KumarAzizChakrabortyLudwigGopalakrishnanPixleyVasseur2024}
A.~Kumar, K.~Aziz, A.~Chakraborty, A.~W.~W. Ludwig,
S.~Gopalakrishnan, J.~H. Pixley, and R.~Vasseur,
``Boundary transfer matrix spectrum of measurement-induced transitions,''
\href{https://doi.org/10.1103/PhysRevB.109.014303}
{\emph{Phys. Rev. B} \textbf{109}, 014303 (2024)}.

\bibitem{VerstraetePoppCirac2004}
F.~Verstraete, M.~Popp, and J.~I. Cirac,
``Entanglement versus correlations in spin systems,''
\href{https://doi.org/10.1103/PhysRevLett.92.027901}
{\emph{Phys. Rev. Lett.} \textbf{92}, 027901 (2004)}.

\bibitem{LinYeZouSangHsieh2023}
C.-J. Lin, W.~Ye, Y.~Zou, S.~Sang, and T.~H. Hsieh,
``Probing sign structure using measurement-induced entanglement,''
\href{https://doi.org/10.22331/q-2023-02-02-910}
{\emph{Quantum} \textbf{7}, 910 (2023)}.

\bibitem{ChengWenGopalakrishnanVasseurPotter2024}
Z.~Cheng, R.~Wen, S.~Gopalakrishnan, R.~Vasseur, and A.~C. Potter,
``Universal structure of measurement-induced information in many-body ground states,''
\href{https://doi.org/10.1103/PhysRevB.109.195128}
{\emph{Phys. Rev. B} \textbf{109}, 195128 (2024)}.

\bibitem{Rajabpour2015}
M.~A. Rajabpour,
``Post-measurement bipartite entanglement entropy in conformal field theories,''
\href{https://doi.org/10.1103/PhysRevB.92.075108}
{\emph{Phys. Rev. B} \textbf{92}, 075108 (2015)}.

\bibitem{Rajabpour2016}
M.~A. Rajabpour,
``Entanglement entropy after a partial projective measurement in \(1+1\)-dimensional conformal field theories: exact results,''
\href{https://doi.org/10.1088/1742-5468/2016/06/063109}
{\emph{J. Stat. Mech.: Theory Exp.} \textbf{2016}, 063109 (2016)}.

\bibitem{NajafiRajabpour2016}
K.~Najafi and M.~A. Rajabpour,
``Entanglement entropy after selective measurements in quantum chains,''
\href{https://doi.org/10.1007/JHEP12(2016)124}
{\emph{J. High Energy Phys.} \textbf{2016}, 124 (2016)}.

\bibitem{WeinsteinSajithAltmanGarratt2023}
Z.~Weinstein, R.~Sajith, E.~Altman, and S.~J. Garratt,
``Nonlocality and entanglement in measured critical quantum Ising chains,''
\href{https://doi.org/10.1103/PhysRevB.107.245132}
{\emph{Phys. Rev. B} \textbf{107}, 245132 (2023)}.

\bibitem{YangMaoJian2023}
Z.~Yang, D.~Mao, and C.-M. Jian,
``Entanglement in a one-dimensional critical state after measurements,''
\href{https://doi.org/10.1103/PhysRevB.108.165120}
{\emph{Phys. Rev. B} \textbf{108}, 165120 (2023)}.

\bibitem{GarrattWeinsteinAltman2023}
S.~J.~Garratt, Z.~Weinstein, and E.~Altman,
``Measurements conspire nonlocally to restructure critical quantum states,''
\href{https://doi.org/10.1103/PhysRevX.13.021026}
{\emph{Phys. Rev. X} \textbf{13}, 021026 (2023)}.

\bibitem{MurcianoSalaLiu2023}
S.~Murciano, P.~Sala, Y.~Liu, R.~S.~K.~Mong, and J.~Alicea,
``Measurement-altered Ising quantum criticality,''
\href{https://doi.org/10.1103/PhysRevX.13.041042}
{\emph{Phys. Rev. X} \textbf{13}, 041042 (2023)}.

\bibitem{LiuMurcianoMrossAlicea2025}
Y.~Liu, S.~Murciano, D.~F.~Mross, and J.~Alicea,
``Boundary transitions from a single round of measurements on gapless quantum states,''
\href{https://doi.org/10.1103/l4b7-h5cd}
{\emph{Phys. Rev. Research} \textbf{7}, 023293 (2025)}.

\bibitem{TaddiaOrtolaniPalmai2016}
L.~Taddia, F.~Ortolani, and T.~P\'almai,
``R\'enyi entanglement entropies of descendant states in critical systems
with boundaries: conformal field theory and spin chains,''
\href{https://doi.org/10.1088/1742-5468/2016/09/093104}
{\emph{J. Stat. Mech.} 2016, 093104 (2016)}.



\bibitem{Chen:2023gql}
M.~Chen and H.-H. Chen,
``R\'enyi entanglement asymmetry in \((1+1)\)-dimensional conformal field theories,''
\href{https://doi.org/10.1103/PhysRevD.109.065009}
{\emph{Phys. Rev. D} \textbf{109}, 065009 (2024)}.

\bibitem{KhannaVasseur2026}
K.~Khanna and R.~Vasseur,
``Measurement-induced entanglement in conformal field theory,''
\href{https://doi.org/10.1103/b7sb-nhjq}
{\emph{Phys. Rev. Lett.} \textbf{136}, 160402 (2026)}.

\bibitem{Zhang:2024rnh}
Y.~Zhang, Y.-H. Wu, L.~Wang, and H.-H. Tu,
``Crosscap states and duality of Ising field theory in two dimensions,''
\href{https://doi.org/10.1103/41qh-vws4}
{\emph{Phys. Rev. B} \textbf{113}, L060412 (2026)}.

\bibitem{WeiYoneta2026}
Z.~Wei and Y.~Yoneta,
``Crosscap quenches and entanglement evolution,''
\href{https://doi.org/10.1007/JHEP05(2026)230}
{\emph{JHEP} \textbf{05}, 230 (2026)}.

\bibitem{Chalas:2024yts}
K.~Chalas, P.~Calabrese, and C.~Rylands,
``Quench dynamics of entanglement from crosscap states,''
\href{https://doi.org/10.21468/SciPostPhys.19.5.132}
{\emph{SciPost Phys.} \textbf{19}, 132 (2025)}.

\bibitem{Chen:2025wzd}
H.-H.~Chen,
``Exact quench dynamics from thermal pure quantum states,''
\href{https://doi.org/10.1103/899k-yxpj}
{\emph{Phys. Rev. B} \textbf{114}, L020302 (2026)}.

\bibitem{Lowdin1955}
P.-O. L{\"o}wdin,
``Quantum theory of many-particle systems. I. Physical interpretations by means of density matrices, natural spin-orbitals, and convergence problems in the method of configurational interaction,''
\href{https://doi.org/10.1103/PhysRev.97.1474}
{\emph{Phys. Rev.} \textbf{97}, 1474--1489 (1955)}.

\bibitem{BalianBrezin1969}
R.~Balian and E.~Br{\'e}zin,
``Nonunitary Bogoliubov transformations and extension of Wick's theorem,''
\href{https://doi.org/10.1007/BF02710281}
{\emph{Nuovo Cimento B} \textbf{64}, 37--55 (1969)}.

\bibitem{Burton2021}
H.~G.~A. Burton,
``Generalized non-orthogonal matrix elements: Unifying Wick's theorem and the Slater--Condon rules,''
\href{https://doi.org/10.1063/5.0045442}
{\emph{J. Chem. Phys.} \textbf{154}, 144109 (2021)}.

\bibitem{Klich2003}
I.~Klich,
``An elementary derivation of Levitov's formula,''
\href{https://doi.org/10.1007/978-94-010-0089-5_19}
{in \emph{Quantum Noise in Mesoscopic Physics},
Y.~V. Nazarov, ed.,
NATO Science Series II, vol.~97, pp.~397--402.
Springer, Dordrecht (2003)}.

\bibitem{Peschel2003}
I.~Peschel,
``Calculation of reduced density matrices from correlation functions,''
\href{https://doi.org/10.1088/0305-4470/36/14/101}
{\emph{J. Phys. A: Math. Gen.} \textbf{36}, L205--L208 (2003)}.

\bibitem{PeschelEisler2009}
I.~Peschel and V.~Eisler,
``Reduced density matrices and entanglement entropy in free lattice models,''
\href{https://doi.org/10.1088/1751-8113/42/50/504003}
{\emph{J. Phys. A: Math. Theor.} \textbf{42}, 504003 (2009)}.

\bibitem{Grover2013}
T.~Grover,
``Entanglement of interacting fermions in quantum Monte Carlo calculations,''
\href{https://doi.org/10.1103/PhysRevLett.111.130402}
{\emph{Phys. Rev. Lett.} \textbf{111}, 130402 (2013)}.

\bibitem{AssaadLangToldin2014}
F.~F. Assaad, T.~C. Lang, and F.~Parisen Toldin,
``Entanglement spectra of interacting fermions in quantum Monte Carlo simulations,''
\href{https://doi.org/10.1103/PhysRevB.89.125121}
{\emph{Phys. Rev. B} \textbf{89}, 125121 (2014)}.

\bibitem{Assaad2015}
F.~F. Assaad,
``Stable quantum Monte Carlo simulations for entanglement spectra of interacting fermions,''
\href{https://doi.org/10.1103/PhysRevB.91.125146}
{\emph{Phys. Rev. B} \textbf{91}, 125146 (2015)}.
\end{thebibliography}
\end{document}